\documentclass[preprint,preprintnumbers,amsmath,amssymb,superscriptaddress]{revtex4}
\usepackage{graphicx,color}
\usepackage{amsmath,amssymb}
\usepackage{url}
\usepackage{epstopdf}
\usepackage{bbold}

\newcommand{\hs}{\hspace*{0.5cm}}

\newcommand{\eq}[1]{Eq.~(\ref{#1})}
\newcommand{\bib}[1]{Ref.~\cite{#1}}

\newcommand{\fig}[1]{Fig.~\ref{#1}}
\newcommand{\tab}[1]{Table~\ref{#1}}
\newcommand{\sect}[1]{Section~\ref{#1}}

\newcommand{\appen}[1]{Appendix~\ref{#1}}

\newcommand{\gev}{{\unskip\,\text{GeV}}}
\newcommand{\tev}{{\unskip\,\text{TeV}}}
\newcommand{\mev}{{\unskip\,\text{MeV}}}
\newcommand{\ev}{{\unskip\,\text{eV}}}

\newcommand{\fb}{{\unskip\,\text{fb}}}

\newcommand{\be}{\begin{equation}}
\newcommand{\ee}{\end{equation}}
\newcommand{\bea}{\begin{eqnarray}}
\newcommand{\eea}{\end{eqnarray}}

\newcommand{\crn}{\nonumber \\}

\newcommand{\la}{\lambda}

\newcommand{\fr}{\frac}
\newcommand{\bc}{\begin{center}}
\newcommand{\ec}{\end{center}}

\newcommand{\ep}{\epsilon}

\newcommand {\ba}{\begin{array}}
\newcommand {\ea}{\end{array}}
\newcommand{\ben}{\begin{enumerate}}
\newcommand{\een}{\end{enumerate}}

\usepackage{epsfig,graphicx}
\usepackage{bm}
\usepackage{dcolumn}

\begin{document}

\preprint{HU-EP-17/33, IFIRSE-TH-2017-1}

\title{Exact one-loop results for $l_i \to l_j\gamma$ in 3-3-1 models}
\author{L.T. Hue}\email{lthue@iop.vast.ac.vn}
\affiliation{\scriptsize
Institute for Research and Development, Duy Tan University, 550000 Da Nang, Vietnam
}
\affiliation{\scriptsize Institute of Physics, Vietnam Academy of Science and Technology, 10 Dao Tan, Ba
Dinh, 100000 Hanoi, Vietnam}
\author{L.D. Ninh}\email{ldninh@ifirse.icise.vn (Corresponding Author)}
\affiliation{\scriptsize
Institute For Interdisciplinary Research in Science and Education, 
ICISE, Ghenh Rang, 590000 Quy Nhon, Vietnam}
\affiliation{\scriptsize
Humboldt-Universit{\"a}t zu Berlin, Institut f{\"u}r Physik, Newtonstrasse 15, 12489 Berlin, Germany}
\author{T.T. Thuc}\email{truongtrongthuck17@gmail.com}
\affiliation{\scriptsize
Department of Education and Training of Ca Mau, 70 Phan Dinh Phung, 970000 Ca Mau, Viet Nam}
\author{N.T.T. Dat}\email{tnguyen4@ictp.it}
\affiliation{\scriptsize
Dipartimento di Fisica, Theoretical section, Universit{\`a} di Trieste,
Strada Costiera 11, 34151 Trieste, Italy}
\affiliation{\scriptsize
International Center for Theoretical Physics, Strada Costiera 11, 34151 Trieste, Italy}

\begin{abstract}
We investigate the decays $l_i\rightarrow l_j \gamma$, with $l_i=e,\mu,\tau$ 
in a general class of 3-3-1 models with heavy exotic leptons with arbitrary electric charges. 
We present full and exact analytical results keeping external lepton masses. 
As a by product, we perform numerical comparisons between exact results and approximate ones where 
the external lepton masses are neglected. As expected, 
we found that branching fractions can reach the current experimental limits if mixings 
and mass differences of the exotic leptons are large enough. We also found unexpectedly that, depending on the 
parameter values, there can be huge destructive interference between the gauge and Higgs contributions when 
the gauge bosons connecting the Standard Model leptons to the exotic leptons are light enough. 
This mechanism should be taken into account when using experimental constraints on the branching fractions 
to exclude the parameter space of the model.  
\end{abstract}

\pacs{{\bf Last updated: \today}
%11.15.Ex  Supersymmetric models,
%12.60.Fr  Extensions of electroweak Higgs sector,
 %13.66.Fg Gauge and Higgs boson production in $e^-\,  e^+$ interactions
 }
\maketitle
%%%%%%%%%%%%%%%%%%%
\section{Introduction}
\label{sec:intro}
The discovery of flavor neutrino oscillations (see \bib{Olive:2016xmw} and the references therein) proves that neutrinos are massive. This leads to an important 
consequence that the lepton-flavor number violating decay $\mu \to e \gamma$ is non-vanishing, being proportional 
to the neutrino masses and the mixing matrix. Assuming tiny neutrino masses satisfying current experimental constraints \cite{Olive:2016xmw}, 
extension of the Standard Model (SM) with right-handed neutrinos 
predicts that the branching ratio is $\text{Br} \approx 10^{-55}$, which will be called 
the SM contribution from now on. 
Meanwhile, the current experimental limits read \cite{Olive:2016xmw}
\begin{align}
&\text{Br}(\mu^- \to e^-\gamma) < 4.2\times 10^{-13},\crn 
&\text{Br}(\tau^- \to e^-\gamma) < 3.3\times 10^{-8},\crn
&\text{Br}(\tau^- \to \mu^-\gamma) < 4.4\times 10^{-8}. 
\end{align}

From theoretical side, the processes $l_i \to l_j \gamma$ are loop induced. 
Given that the SM contribution is strongly suppressed, they can be good places to look for 
new physics. In this paper, we consider a simple extension of the SM using the local 
gauge group of $SU(3)_C\otimes SU(3)_L\otimes U(1)_X$ (3-3-1) with new exotic leptons. 
The word exotic here means that they can have arbitrary electric charges and arbitrary masses.   
In this model, the electron (and similarly for 
muon and tauon) together with a neutrino and a new exotic lepton are in a triplet (or anti-triplet) representation of $SU(3)_L$. 
In this work, we calculate both neutrino and exotic-lepton contributions, with special attention to the latter because the former 
is numerically suppressed as above mentioned. 

We remark that 3-3-1 model is an active field of research and has a long history, see \bib{Singer:1980sw} 
and references therein. In this work, we choose a general class of 3-3-1 models, which are similar to 
the models presented in Refs.~\cite{Singer:1980sw,Pleitez:1992xh,Ozer:1995xi} where new heavy leptons are introduced. However, there is an important 
difference: instead of fixing the electric charges of the new leptons to specific values being $0$, $+1$ or $-1$, we let them be arbitrary. We will then study 
the dependence of the $l_i \to l_j \gamma$ branching fractions on this arbitrary charge. That class of 3-3-1 models has been 
studied in many works, see e.g. \cite{Diaz:2004fs,Buras:2012dp}. If we replace the new leptons with charge-conjugated partners 
of the SM leptons, we will have different 3-3-1 models with lepton-number violation, see e.g. Refs.~\cite{Valle:1983dk,Pisano:1991ee,Foot:1992rh,Frampton:1992wt,Foot:1994ym}. The decays $l_i \to l_j \gamma$ in these models 
have been discussed in Refs.~\cite{Boucenna:2015zwa,Machado:2016jzb}, see also the recent review \cite{Lindner:2016bgg} 
and references therein. We do not discuss these types of models in this work, but rather focus on the case with exotic leptons.    

In the general class of 3-3-1 models here considered, there is one important parameter usually called $\beta$, which together with 
$X$, the new charge corresponding to 
the group $U(1)_X$, define the electric-charge operator. The electric charges of new particles therefore depend on $\beta$. 
It has been known and widely accepted that $\beta$ is one of the most important parameters to classify 3-3-1 models. 

Recently, there occurred new efforts using 3-3-1 models to understand tensions between experimental measurements and the SM 
results in B physics, see e.g. \cite{Buras:2013dea,Buras:2016dxz}. Motivated by this work, we want to use 3-3-1 models to understand 
the $l_i \to l_j \gamma$ decays. Since the new leptons are assumed to be heavy, we expect large branching fractions. 
However, this is not totally obvious, because there are two contributions from gauge and Higgs sectors. Does a destructive interference effect occur?   

The aim of this paper is manifold. First, we calculate the full and exact result for $l_i \to l_j \gamma$ partial decay widths for 
a general class of 3-3-1 models with arbitrary $\beta$. As a by product, we will perform numerical comparisons between the exact results (i.e. external lepton masses are kept) and 
approximate ones where external lepton masses are neglected. We note that approximate results have been almost exclusively used in the literature for the SM and many other models. 
We found this uncomfortable because the neutrino masses, which are much smaller than the 
lepton masses, are kept. We therefore want to know to what accuracy the approximate results valid, using the SM with arbitrary neutrino masses to answer this. 
As far as we know, this important point has never been addressed in the literature. 
We will also perform numerical studies for 3-3-1 model to see whether destructive interference effects occur and to see 
the dependence on $\beta$, gauge boson and Higss masses. To the best of our knowledge, this is the first study of $l_i \to l_j \gamma$ in 3-3-1 models 
with exotic leptons.    

The paper is organized as follows. In the next section, we review the model and calculate the Feynman rules needed for $l_i \to l_j \gamma$ 
decays. We then summarize the main calculation steps and present analytical results in \sect{analytical_results}. 
Numerical results are discussed in \sect{numerical_results}. In \sect{results_neutrino} we perform comparisons between the approximate and exact results for the neutrino contribution. 
In \sect{results_exotic} we present results for the exotic-lepton contribution. Conclusions are in \sect{conclusion}. 
Finally, we provide two Appendices \ref{appen_loops} and \ref{appen_approx} to complete the results of \sect{analytical_results}.

%%%
\section{3-3-1 model with arbitrary $\beta$}
\label{models}
One important condition we require is that the 3-3-1 model has to match the SM at the energy of the EW scale, about $250\gev$. 
This means that the $SU(3)_L$ symmetry is valid at a higher energy scale and is spontaneously broken down to the $SU(2)_L$ symmetry using the Brout-Englert-Higgs mechanism. In order to match the fermion representation of the SM, the simplest choice is to assign fermions into triplets and anti-triplets of the $SU(3)_L$ group. However, this requires new fermions. In general, the electric charges of these new fermions are unkown. They, however, cannot be totally arbitrary because of the symmetry and of the matching condition with the SM. In most general terms, the electric charge operator can be written as
\bea
Q = T_3 + \beta T_8 + X \mathbb{1},
\label{eq:charge_Q}
\eea
where we have introduced the $SU(3)$ generators $T_3$, $T_8$. 
Thus, the charge operator $Q$ depends on two parameters $\beta$ and $X$. With this information, we can write down 
the lepton representation as follows. Left-handed leptons are assigned to anti-triplets and right-handed leptons to 
singlets: 
\bea && L'_{aL}=\left(
       \begin{array}{c}
         e'_a \\
         -\nu'_{a} \\
         E'_a \\
       \end{array}
     \right)_L \sim \left(3^*~, -\frac{1}{2}+\frac{\beta}{2\sqrt{3}}\right), \hs a=1,2,3,\crn
     && e'_{aR}\sim   \left(1~, -1\right)  , \hs \nu'_{aR}\sim  \left(~1~, 0\right) ,\hs E'_{aR} \sim   \left(~1~, -\frac{1}{2}+\frac{\sqrt{3}\beta}{2}\right).  \label{lep}\eea
  The model includes three RH neutrinos $\nu'_{aR}$ and exotic leptons $E'^a_{L,R}$ which are much heavier than the normal leptons. 
  The prime denotes flavor states to be distinguished with mass eigenstates introduced later. 
  The numbers in the parentheses are to label the representation of $SU(3)_L\otimes U(1)_X$ group. 
  For singlets, we have $Q=X$ and hence the electric charges of the new leptons can be read off from the above information. 
The quark sector is not here specified since it is irrelevant to our present work.  

We now discuss gauge and Higgs interactions. 
There are totally 9 EW gauge bosons, included in the following covariant derivative
 \bea D_{\mu}\equiv \partial_{\mu}-i g T^a W^a_{\mu}-i g_X X T^9X_{\mu},  \label{coderivative1}\eea
where $T^9=\mathbb{1}/\sqrt{6}$, $g$ and $g_X$ are coupling constants corresponding to the two groups $SU(3)_L$ and $U(1)_X$, respectively. 
The matrix $W^aT^a$, where $T^a =\lambda_a/2$ corresponding to a triplet representation, can be written as
 \bea W^a_{\mu}T^a=\frac{1}{2}\left(
                     \begin{array}{ccc}
                       W^3_{\mu}+\frac{1}{\sqrt{3}} W^8_{\mu}& \sqrt{2}W^+_{\mu} &  \sqrt{2}Y^{+A}_{\mu} \\
                        \sqrt{2}W^-_{\mu} &  -W^3_{\mu}+\frac{1}{\sqrt{3}} W^8_{\mu} & \sqrt{2}V^{+B}_{\mu} \\
                       \sqrt{2}Y^{-A}_{\mu}& \sqrt{2}V^{-B}_{\mu} &-\frac{2}{\sqrt{3}} W^8_{\mu}\\
                     \end{array}
                   \right),
  \label{wata}\eea
where we have defined the mass eigenstates of the charged gauge bosons as
\bea W^{\pm}_{\mu}=\frac{1}{\sqrt{2}}\left( W^1_{\mu}\mp i W^2_{\mu}\right),\crn
Y^{\pm A}_{\mu}=\frac{1}{\sqrt{2}}\left( W^4_{\mu}\mp i W^5_{\mu}\right),\crn
V^{\pm B}_{\mu}=\frac{1}{\sqrt{2}}\left( W^6_{\mu}\mp i W^7_{\mu}\right).
   \label{gbos}\eea 
From \eq{eq:charge_Q}, the electric charges of the gauge bosons are calculated as 
\bea
A=\fr{1}{2}+\beta\fr{\sqrt{3}}{2}, \quad 
B=-\fr{1}{2}+\beta\fr{\sqrt{3}}{2}\label{charge_AB}.\eea 
We note that $B$ is also the electric charge of the new leptons $E_a$.

To generate masses for gauge bosons and fermions, we need three scalar triplets. They are 
defined as
  \bea && \chi=\left(
              \begin{array}{c}
                \chi^{+A} \\
                \chi^{+B} \\
                \chi^0 \\
              \end{array}
            \right)\sim \left(3~, \frac{\beta}{\sqrt{3}}\right), \hs  \rho=\left(
              \begin{array}{c}
                \rho^+ \\
                \rho^0 \\
                \rho^{-B} \\
              \end{array}
            \right)\sim \left(3~, \frac{1}{2}-\frac{\beta}{2\sqrt{3}}\right)\crn
  && \eta=\left(
              \begin{array}{c}
                \eta^0 \\
                \eta^- \\
                \eta^{-A} \\
              \end{array}
            \right)\sim \left(3~, -\frac{1}{2}-\frac{\beta}{2\sqrt{3}}\right),
    \label{higgsc}
  \eea
where $A,B$ denote electric charges as defined in \eq{charge_AB}. 
These Higgses develop vacuum expectation values (VEV) defined as
  \bea&& \langle  \chi\rangle=\frac{1}{\sqrt{2}}\left(
              \begin{array}{c}
                0 \\
                0 \\
                u \\
              \end{array}
            \right), \hs \langle  \rho \rangle =\frac{1}{\sqrt{2}}\left(
              \begin{array}{c}
                0 \\
                v \\
                0 \\
              \end{array}
            \right),\hs \langle   \eta \rangle= \frac{1}{\sqrt{2}}\left(
              \begin{array}{c}
                v' \\
                0 \\
                0 \\
              \end{array}
            \right). \label{vevhigg1}\eea
 The symmetry breaking happens in two steps:  
$SU(3)_L\otimes U(1)_X\xrightarrow{u} SU(2)_L\otimes U(1)_Y\xrightarrow{v,v'} U(1)_Q$.  
It is therefore reasonable to assume that $u > v,v'$. 
After the first step, five gauge bosons will be massive and the remaining four massless gauge bosons 
can be identified with the before-symmetry-breaking SM gauge bosons. This leads to the following matching condition 
for the couplings
\bea
g_2 = g,\quad g_1 = g_X\fr{g}{\sqrt{6g^2 + \beta^2 g_X^2}},
\label{matching_coupl} 
\eea
where $g_2$ and $g_1$ are the two couplings of the SM corresponding to $SU(2)_L$ and $U(1)_Y$, respectively. 
From this we get the following important equation which helps to constrain $\beta$: 
\bea
\fr{g_X^2}{g^2} = \fr{6s_W^2}{1-(1+\beta^2)s_W^2},
\label{eq_beta_coupl}
\eea
where the weak mixing angle is defined as $t_W = \tan\theta_W = g_1/g_2$ and we denote $s_W = \sin\theta_W$.  
Putting in the value of $s_W$, we get approximately
\bea
|\beta| \le \sqrt{3},
\label{eq_beta_constraint}
\eea
which will be used in the numerical analysis. 

The masses of the charged gauge bosons are
\bea 
m^2_{Y^{\pm A}} = \frac{g^2}{4}(u^2+v^{\prime 2}),\quad 
m^2_{V^{\pm B}}=\frac{g^2}{4}(u^2+v^2),\quad
m^2_{W^\pm} = \fr{g^2}{4}(v^2 + v^{\prime 2}).
\label{masga}\eea 

We now discuss the mixings of leptons. In general, the mixing between a SM lepton and a new lepton is allowed 
if they have the same electric charge. However, since we consider a general class of models with arbitrary $\beta$, this 
mixing effect will be neglected. This is justified because we will assume that the new leptons are much heavier than the SM leptons. Therefore, only generation mixings as in the SM are allowed. The Yukawa Lagrangian related to these mixings reads
\bea \mathcal{L}^\text{yuk}_{\mathrm{lepton}}= -Y^e_{ab} \overline{L'}_{aL} \eta^*e'_{bR}- Y^\nu_{ab} \overline{L'}_{aL} \rho^*\nu'_{bR} - Y^E_{ab} \overline{L'}_{aL} \chi^*E'_{bR}+\mathrm{h.c.}, \label{ylepton1}\eea
where $a,b=e,\mu,\tau$ are family indices. The corresponding mass terms are:
\bea \mathcal{L}^\text{mass}_{\mathrm{lepton}}= -\frac{Y^e_{ab}v'}{\sqrt{2}} \overline{e'}_{aL}e'_{bR}+\frac{Y^\nu_{ab}v}{\sqrt{2}} \overline{\nu'}_{aL} \nu'_{bR} - \frac{Y^E_{ab}u}{\sqrt{2}} \overline{E'}_{aL} E'_{bR}+\mathrm{h.c.}. \label{mlepton1}\eea
%---

From now on we will work in the basis where the SM charged leptons are in their mass eigenstates. This can always be done without loss of generality. We 
can therefore set $Y^e_{ab}$ to be diagonal and $e' = e$ in Eqs.~(\ref{ylepton1},\ref{mlepton1}). The transformations from the flavor states to mass eigenstates are defined as
\bea
&&\nu'_{aL} = U^L_{ab} \nu_{bL}, \quad \nu'_{aR} = U^R_{ab} \nu_{bR},\crn
&&E'_{aL} = V^L_{ab} E_{bL}, \quad E'_{aR} = V^R_{ab} E_{bR}, 
\label{mixing_E_LR}
\eea 
where $U^{L,R}$ and $V^{L,R}$ are $3\times 3$ unitary matrices for the neutrinos and new leptons, respectively. 
The matrix $V^L$, included in the vertices of the SM charged leptons and the new leptons, 
is similar to the matrix $U^L=U_\text{PMNS}$. 

For the Higgs sector, we assume $A,B \neq 0, \pm 1$, so that only the following mixings of scalar fields with the same 
electric charge are allowed, namely
$(\chi_{A},~\eta_{A})$, $(\chi_{B},~\rho_{B})$, $(\rho^+,\eta^+)$ and $(\chi^0,~\rho^0,~ \eta^0)$. 
The neutral components are expanded as:
   %---
   \bea &&\chi^0=\frac{1}{\sqrt{2}} \left(u+\xi_{\chi}+i \zeta_{\chi}\right), \hs \langle\xi_{\chi}\rangle=\langle\zeta_{\chi}\rangle=0,
   \crn &&  \rho^0=\frac{1}{\sqrt{2}} \left(v+\xi_{\rho}+i \zeta_{\rho}\right), \hs \langle\xi_{\rho}\rangle=\langle\zeta_{\rho}\rangle=0,
   \crn &&   \eta^0=\frac{1}{\sqrt{2}} \left(v'+\xi_{\eta}+i \zeta_{\eta}\right), \hs \langle\xi_{\eta}\rangle=\langle\zeta_{\eta}\rangle=0.
   \label{neuHiggs}\eea
The ratios between  VEVs are used to define three mixing angles:
 \be s^2_{v'v}=\sin^2\beta_{v'v}= \frac{v'^2}{v^2+v'^2}, \;  s^2_{vu}=\sin^2\beta_{vu}= \frac{v^2}{u^2+v^2}, \; 
s^2_{v'u}=\sin^2\beta_{v'u}= \frac{v'^2}{v'^2+u^2}. \label{vevang}\ee
We will also use the following notation $t_{v'v}=s_{v'v}/c_{v'v}$, $t_{v'u}=s_{v'u}/c_{v'u}$.\\
The scalar potential is
\bea V_{\mathrm{h}}&=&\mu_1^2 \eta^{\dagger}\eta+\mu_2^2\rho^{\dagger}\rho+\mu_3^2\chi^{\dagger}\chi
+\lambda_1 \left(\eta^{\dagger}\eta\right)^2
+\lambda_2\left(\rho^{\dagger}\rho\right)^2
+\lambda_3\left(\chi^{\dagger}\chi\right)^2\crn
&+& \lambda_{12}(\eta^{\dagger}\eta)(\rho^{\dagger}\rho)
+\lambda_{13}(\eta^{\dagger}\eta)(\chi^{\dagger}\chi)
+\lambda_{23}(\rho^{\dagger}\rho)(\chi^{\dagger}\chi)\crn
&+&\tilde{\lambda}_{12} (\eta^{\dagger}\rho)(\rho^{\dagger}\eta) 
+\tilde{\lambda}_{13} (\eta^{\dagger}\chi)(\chi^{\dagger}\eta)
+\tilde{\lambda}_{23} (\rho^{\dagger}\chi)(\chi^{\dagger}\rho)\crn
&+&\sqrt{2} f\left(\epsilon_{ijk}\eta^i\rho^j\chi^k +\mathrm{h.c.} \right).\label{hpo1}\eea
With the above notation, the mass eigenstates are
 \bea \left(
          \begin{array}{c}
            \phi^{\pm}_W \\
            H^{\pm} \\
          \end{array}
        \right)=\left(
                  \begin{array}{cc}
                    c_{v'v} & -s_{v'v} \\
                    s_{v'v} & c_{v'v} \\
                  \end{array}
                \right)\left(
                         \begin{array}{c}
                           \rho^{\pm} \\
                           \eta^{\pm} \\
                         \end{array}
                       \right),
   \label{scHigg}\\
   \left(
          \begin{array}{c}
            \phi^{ \pm A}_Y \\
            H^{\pm A} \\
          \end{array}
        \right)=\left(
                  \begin{array}{cc}
                    s_{v'u} & -c_{v'u} \\
                    c_{v'u} & s_{v'u} \\
                  \end{array}
                \right)\left(
                         \begin{array}{c}
                           \eta^{\pm A} \\
                           \chi^{\pm A} \\
                         \end{array}
                       \right),
   \label{qacHigg}\\
   \left(
          \begin{array}{c}
            \phi^{\pm B}_V \\
            H^{\pm B} \\
          \end{array}
        \right)=\left(
                  \begin{array}{cc}
                    s_{vu} & -c_{vu} \\
                    c_{vu} & s_{vu} \\
                  \end{array}
                \right)\left(
                         \begin{array}{c}
                           \rho^{\pm B} \\
                           \chi^{\pm B} \\
                         \end{array}
                       \right),
   \label{qbcHigg}   
   \eea
   where 
   $\phi_W^\pm$, $\phi^{ \pm A}_Y$ and $\phi^{\pm B}_V$  are the Goldstone bosons of $W^\pm$, $Y^{\pm A}$ and $V^{\pm B}$, respectively. The masses of the charged Higgs bosons are
   \bea
   &&m^2_{H^{\pm}}=(v^2 + v^{\prime 2})\left(\frac{-fu}{v'v}+\frac{1}{2}\tilde{\lambda}_{12}\right),\crn
   &&m^2_{H^{\pm A}}=(u^2+v^{\prime 2})\left(\frac{-fv}{v'u}+\frac{1}{2}\tilde{\lambda}_{13}\right),\crn
   &&m^2_{H^{\pm B}}=(u^2+v^2)\left(\frac{-fv'}{uv}+\frac{1}{2}\tilde{\lambda}_{23}\right).\label{charge_Higgs_mass}
   \eea
The neutral Higgs bosons are not involved in our calculation; hence they have been ignored. In total, there are six charged Higgs bosons, one neutral pseudoscalar Higgs and three neutral scalar Higgses. Bosonic particles with electric charges of $\pm B$ are not involved in the present calculation. 
Nevertheless, their masses and mixing angles are above provided for the sake of completeness.    
 
From the above information we can obtain all vertices needed for the calculation of $l_i \to l_j \gamma$ decays. They are listed in \tab{hl331table}.
\begin{table}[h]
\begin{tabular}{|c|c|c|c|}
\hline
Vertex & Coupling & Vertex & Coupling \\
\hline
$\overline{\nu}_{a}e_{b}H^+$ & $ \fr{ig}{\sqrt{2}m_W}U^{L*}_{ba} \left( \frac{m_{e_b}}{t_{v'v}} P_R+m_{\nu_a}t_{v'v} P_L\right) $ &
$\overline{e_a}\nu_bH^-$ & $ \fr{ig}{\sqrt{2}m_W}U^{L}_{ab}\left( \frac{m_{e_a}}{t_{v'v}} P_L+ m_{\nu_b}t_{v'v}  P_R\right) $ \\
\hline
$\overline{E}_ae_b H^{+A}$&$\fr{-ig}{\sqrt{2}m_Y}V^{L*}_{ba}\left( \frac{m_{e_b}}{t_{v'u}} P_R+ m_{E_a} t_{v'u} P_L\right)$
&$\overline{e}_aE_bH^{-A} $&$\fr{-ig}{\sqrt{2}m_Y}V^L_{ab}\left( \frac{m_{e_a}}{t_{v'u}}  P_L+ m_{E_b} t_{v'u}  P_R\right)$\\
\hline
$\overline{\nu}_{a}e_{b}W^{+\mu}$ &$\frac{ig}{\sqrt{2}}U^{L*}_{ba}\gamma_\mu P_L $ &$\overline{e}_a\nu _b W^{-\mu}$
 & $\frac{ig}{\sqrt{2}}U^{L}_{ab}\gamma_\mu P_L $\\
\hline
$\overline{\nu}_{a}e_{b}\phi_{W}^{+}$ &$\frac{-ig}{\sqrt{2}m_W}U^{L*}_{ba}(m_{e_b}P_R - m_{\nu_a} P_L) $ &$\overline{e}_a\nu_b \phi_{W}^{-}$
 & $\frac{-ig}{\sqrt{2}m_W}U^{L}_{ab}(m_{e_a} P_L - m_{\nu_b} P_R) $\\
\hline
$\overline{E}_ae_b Y^{+A\mu}$ &$\frac{-ig}{\sqrt{2}}V^{L*}_{ba}\gamma_\mu P_L$ &$\overline{e}_aE_bY^{-A\mu} $
 & $\frac{-ig}{\sqrt{2}}V^L_{ab}\gamma_\mu P_L$\\ 
\hline
$\overline{E}_ae_b \phi_{Y}^{+A}$ &$\frac{-ig}{\sqrt{2} M_Y}V^{L*}_{ba}(m_{e_b}P_R - m_{E_a}P_L)$ &$\overline{e}_aE_b\phi_{Y}^{-A} $
 & $\frac{-ig}{\sqrt{2}M_Y}V^L_{ab}(m_{e_a}P_L - m_{E_b}P_R)$\\ 
\hline
 $A^{\la}W^{+\mu}W^{-\nu}$&$-ie\Gamma_{\la \mu \nu}(p_A,p_{W^+},p_{W^-}) $&$A^{\la}Y^{+A\mu}Y^{-A\nu}$&$ -ieA \Gamma_{\la\mu\nu}(p_A,p_{Y^{+A}},p_{Y^{-A}})$\\
\hline
 $A^{\la}W^{\pm\mu}\phi_W^{\mp}$ & $iem_Wg_{\la\mu}$ & $A^{\la}Y^{\pm A\mu}\phi_Y^{\mp A}$ & $-ieAm_Y g_{\la\mu}$\\
\hline
 $A^{\mu}H^+H^-$&$ie(p_{H^+}-p_{H^-})_\mu$&$A^{\mu}H^{+A}H^{-A}$&$ieA(p_{H^{+A}}-p_{H^{-A}})_\mu$\\
\hline
 $A^{\mu}\phi_W^+ \phi_W^-$&$ie(p_{\phi_W^+}-p_{\phi_W^-})_\mu$&$A^{\mu}\phi_Y^{+A}\phi_Y^{-A}$&$ieA(p_{\phi_Y^{+A}}-p_{\phi_Y^{-A}})_\mu$\\
\hline
$A^{\mu}\bar{l}_al_a$&$-ie\gamma_\mu $&$A^{\mu}\overline{E}_a E_a$&$ieB\gamma_\mu  $\\
\hline
\end{tabular}
\caption{Vertices and couplings for $l_i\rightarrow l_j\gamma$ decays in the 3-3-1 model with arbitrary $\beta$ and new leptons. 
All momenta are defined as incoming. The photon field is denoted as $A^\mu$, $a,b = 1,2,3$ are family indices and $\Gamma_{\la \mu \nu}(p_1,p_2,p_3)=(p_1-p_2)_\nu g_{ \la\mu} + (p_2-p_3)_\la g_{\mu\nu}+(p_3-p_1)_\mu g_{ \nu\la}$. 
Other notations are defined in the text. 
\label{hl331table}}
\end{table}

\section{Analytical results}
\label{analytical_results}
Equipped with the above Feynman rules, we can proceed to 
calculate the partial decay width of $l_1 \to l_2 \gamma$ using standard techniques 
of one-loop calculation. We have done this in a careful way, with at least two independent calculations, 
and paid special attention to the relative sign between the gauge and Higgs contributions. 
This relative sign is very important because, as we will see in the numerical results, the interference 
term can be positive or negative.  

In the literature, the calculation of $l_1 \to l_2 \gamma$ is usually done by neglecting the external 
lepton masses. As stated in the introduction, we found this uneasy because the neutrino masses, which are much smaller than the 
lepton masses, are kept. We therefore want to check the validity of this approximation. 
To achieve this we have to keep the external lepton masses.  

We have calculated the partial decay width of $l_1 \to l_2 \gamma$ from scratch 
without approximation. In the following we summarize the key points 
and present exact analytical results. Results for the SM case are obtained as 
a special case and are discussed in \sect{results_neutrino}.  

We consider the process 
\bea
l_1(p_1) \to l_2(p_2) + \gamma(q),
\label{eq:proc_mu_e_gamma}
\eea
where $p_1 = p_2+q$ and the helicity indices have been omitted for simplicity. 
The amplitude reads
\bea
\mathcal{M} = \ep^\lambda (q) \bar{u}_1(p_1)\Gamma_\lambda u_2(p_2),
\label{eq:amp} 
\eea 
where $\ep^\lambda$ is the photon's polarization vector, $\Gamma_\lambda$ are $4\times 4$ matrices depending on the gamma matrices, external momenta and coupling constants. 
After requiring the general conditions that the spinors obey the Dirac equations, $q^\mu \ep_\mu =0$, and $q^\lambda \bar{u}_1(p_1)\Gamma_\lambda u_2(p_2) = 0$, we can prove that 
the amplitude depends on only two form factors as
\bea
\mathcal{M} &=& 2(p_1\cdot\epsilon)
\left[C_L\bar{u}_2(p_2)P_Lu_1(p_1)+C_R\bar{u}_2(p_2)P_Ru_1(p_1)\right]\crn
 &-& (m_1C_R + m_2C_L) \bar{u}_2(p_2)/\!\!\!\epsilon P_L u_1(p_1) - (m_1C_L + m_2C_R)\bar{u}_2(p_2)/\!\!\!\epsilon P_R u_1(p_1), 
\label{Amlfv}\eea
where $C_{L,R}$ are called form factors, $P_L = (1-\gamma_5)/2$, $P_R = (1+\gamma_5)/2$. 
The partial decay width is then written as
\be 
\Gamma(l_1\rightarrow l_2\gamma)= \frac{(m^2_1-m^2_2)^3}{16\pi m^3_1}\left(|C_L|^2+|C_R|^2\right).\label{dewidth1}\ee
This result is well known and has been given in e.g. \bib{Lavoura:2003xp}.

Since we assume that the exotic leptons are much heavier than the SM leptons, the branching fractions of the dominant decays of $l_1 \to l_2 \bar{\nu}_2 \nu_1$ in the 3-3-1 model are the same as those of the SM. Using the well-known tree-level result of $\Gamma(l_1 \to l_2 \bar{\nu}_2 \nu_1) = G_F^2 m_1^5/(192\pi^3)$ (see e.g. \cite{Griffiths:2008zz}), where $G_F$ is the Fermi coupling constant, we write the branching fraction as
\bea
\text{Br}(l_1 \to l_2 \gamma) = \fr{12\pi^2}{G_F^2}\left(|D_L|^2+|D_R|^2\right)\text{Br}(l_1 \to l_2 \bar{\nu}_2 \nu_1),
\label{eq:branching_fraction}
\eea 
where $G_F=g^2/(4\sqrt{2}m^2_W)$ and we have defined $C_{L,R} = m_1 D_{L,R}$ and 
the approximation $m_2 \ll m_1$ has been used for the first factor, but not for $D_{L,R}$. 
For later numerical analysis we will use $\text{Br}(\mu \to e \bar{\nu}_e \nu_\mu) = 100\%$, $\text{Br}(\tau \to e \bar{\nu}_e \nu_\tau) = 17.82\%$ and $\text{Br}(\tau \to \mu \bar{\nu}_\mu \nu_\tau) = 17.39\%$ as given in \bib{Olive:2016xmw}. 
It is noted that $D_L\propto \mathcal{O}(m_2/m_1)$ (since only left-handed electron can participate in $SU(3)_L$ interactions) 
and $D_R\propto \mathcal{O}(1)$, and hence, in the approximation $m_2 \ll m_1$, we have $\text{Br}(l_1 \to l_2 \gamma) \propto |D_R|^2$. 
This point is important to understand the approximate results discussed in the next sections. 

\begin{figure}[h]
  \centering
  % Requires \usepackage{graphicx}
  \includegraphics[width=11.0cm]{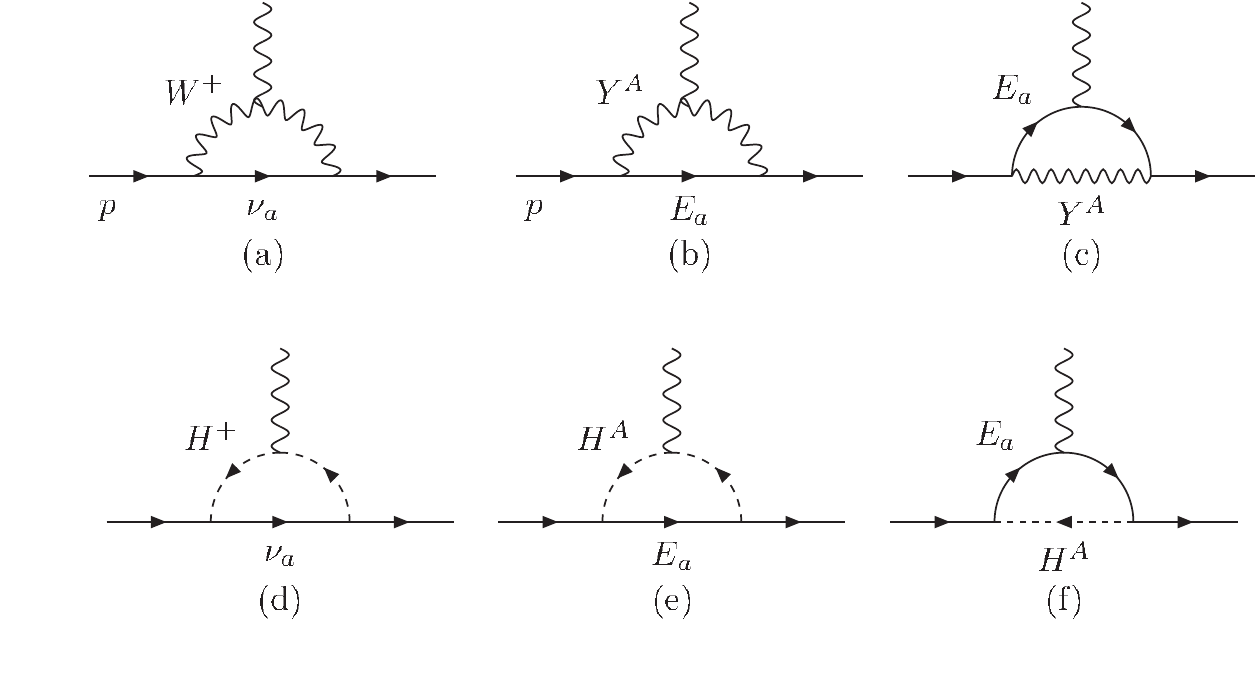}
  \caption{Representative Feynman diagrams contributing to $l_1\rightarrow l_2 \gamma$ decays. 
  There are two groups: the neutrino contribution (a,d) and the exotic lepton contribution (b,c,e,f). 
  \label{dclepton2}}
\end{figure}

The next step is to calculate $D_{L,R}$ for the 3-3-1 model with arbitrary beta presented in the previous section. 
Representative Feynman diagrams are shown in \fig{dclepton2}.  
%---begin: plots as functions of m_{nu1}---%
%---end: plots as functions of m_{nu1}---%
Using the Feynman rules in \tab{hl331table} and 
summing over all possible Feynman diagrams, we obtain the following results
\bea
D_{L,R}=D_{L,R}^{\nu W}+D_{L,R}^{\nu H^+}+D_{L,R}^{E Y}+D_{L,R}^{E H^A},
\label{eq:sum_analytical_results}
\eea
where
{\small
\begin{align} 
D_R^{\nu W} &= -\fr{ieg^2}{32\pi^{2}m_W^2}\sum_{a=1}^3 U_{i_1 a}^{L \star}U_{i_2 a}^{L}\left( 2m^2_W g_1^{\nu_a WW} + m^2_{\nu_a}g_2^{\nu_a WW} + m^2_2g_3^{\nu_a WW} \right),\crn
D_L^{\nu W} &= -\fr{ieg^2m_2}{32\pi^{2}m_W^2m_1}\sum_{a=1}^3 U_{i_1 a}^{L \star}U_{i_2 a}^{L}\left( 2m_W^2g_4^{\nu_a WW}
+m^2_{\nu_a}g_5^{\nu_a WW}
+m_1^2g_6^{\nu_a WW} \right),\crn
D_R^{\nu H^+} &= -\fr{ieg^2}{32\pi^{2}m_W^2}\sum_{a=1}^3 U_{i_1 a}^{L \star}U_{i_2 a}^{L}\left( m^2_{\nu_a}t^2_{v'v}h_1^{\nu_a H^+H^+} + \fr{m^2_2}{t^2_{v'v}}h_2^{\nu_a H^+H^+} + m^2_{\nu_a}h_3^{\nu_a H^+H^+} \right),\crn
D_L^{\nu H^+} &= -\fr{ieg^2m_2}{32\pi^{2}m_W^2m_1}\sum_{a=1}^3 U_{i_1 a}^{L \star}U_{i_2 a}^{L}\left(\fr{m^2_1}{t^2_{v'v}}h_1^{\nu_a H^+H^+} 
+ m^2_{\nu_a}t^2_{v'v}h_2^{\nu_a H^+H^+}
+ m^2_{\nu_a}h_3^{\nu_a H^+H^+}
\right),\crn
D_R^{E Y} &= -\fr{ieg^2}{32\pi^{2}m_Y^2}\sum_{a=1}^3 V_{i_1 a}^{L \star}V_{i_2 a}^{L}\left[A\left( 2m^2_Y g_1^{E_a YY} + m^2_{E_a}g_2^{E_a YY} + m^2_2g_3^{E_a YY} \right)\right. \crn
&+\left. B\left(2m^2_Y g_7^{YE_aE_a} + m^2_{E_a} g_8^{YE_aE_a} + m^2_2 g_9^{YE_aE_a}\right)  \right],\crn
D_L^{E Y} &= -\fr{ieg^2m_2}{32\pi^{2}m_Y^2m_1}\sum_{a=1}^3 V_{i_1 a}^{L \star}V_{i_2 a}^{L}\left[A\left( 2m_Y^2g_4^{E_a YY}
+m^2_{E_a}g_5^{E_a YY}
+m_1^2g_6^{E_a YY} \right)\right.\crn
&+\left. B\left(2m^2_Y g_{10}^{YE_aE_a}+m^2_{E_a} g_{11}^{YE_aE_a}+m^2_1 g_{12}^{YE_aE_a}\right)  \right],\crn
D_R^{E H^A} &= -\fr{ieg^2}{32\pi^{2}m_Y^2}\sum_{a=1}^3 V_{i_1 a}^{L \star}V_{i_2 a}^{L}\left[A\left( m^2_{E_a}t^2_{v'u}h_1^{E_a H^AH^A} + \fr{m^2_2}{t^2_{v'u}}h_2^{E_a H^AH^A} + m^2_{E_a}h_3^{E_a H^AH^A} \right)\right.\crn
&+\left. B\left( m^2_{E_a}t^2_{v'u}h_4^{H^AE_aE_a} + \fr{m^2_2}{t^2_{v'u}}h_5^{H^AE_aE_a} + m^2_{E_a}h_6^{H^AE_aE_a} \right)  \right],\crn
D_L^{E H^A} &= -\fr{ieg^2m_2}{32\pi^{2}m_Y^2m_1}\sum_{a=1}^3 V_{i_1 a}^{L \star}V_{i_2 a}^{L}\left[A\left(\fr{m^2_1}{t^2_{v'u}}h_1^{E_a H^AH^A} 
+ m^2_{E_a}t^2_{v'u}h_2^{E_a H^AH^A}
+ m^2_{E_a}h_3^{E_a H^AH^A}
\right)\right.\crn
&+\left. B\left( \fr{m^2_1}{t^2_{v'u}}h_4^{H^AE_aE_a} 
+ m^2_{E_a}t^2_{v'u}h_5^{H^AE_aE_a}
+ m^2_{E_a}h_6^{H^AE_aE_a} \right)  \right],
\label{eq:list_analytical_results}
\end{align}
} where the loop functions $h_i$ and $g_i$ are simple linear combinations of Passarino-Veltman one-loop 3-point functions as 
given in \appen{appen_loops}. The above writing is inspired by Lavoura \cite{Lavoura:2003xp}. 
Our results have been checked by three different calculations using (i) unitary gauge, (ii) 't Hooft-Feynman gauge, 
and (iii) general formulas of \bib{Lavoura:2003xp}. 
We have classified the results into neutrino and exotic-lepton groups. Each of 
these groups includes Higgs and gauge contributions. In the 't Hooft-Feynman gauge, the gauge contribution includes gauge-gauge, Goldstone-gauge and Goldstone-Goldstone diagrams. We have used FORM \cite{Vermaseren:2000nd,Kuipers:2012rf} to calculate the amplitudes. 

The results can be further simplified if $m_{E_a} \ll m_Y$ and $m_{E_a} \ll m_{H^A}$ with $a=1,2,3$ as 
presented in \appen{appen_approx}. 

Finally, we make an important remark on the dependence on coupling constants. 
From \eq{eq:list_analytical_results} we have $D_{L,R}\propto eg^2$. Using 
\eq{eq:branching_fraction} and noticing that $G_F=g^2/(4\sqrt{2}m^2_W)$, we get 
$\text{Br}(l_1 \to l_2 \gamma) \propto e^2$, being independent of $g$ or $s_W$. Clearly, 
the coupling constant $e=\sqrt{4\pi\alpha}$ should be calculated in the low-energy limit 
for the processes at hand. Therefore, we will use $\alpha(0)$ as input parameter in 
our numerical analyses. 

\section{Numerical results}
\label{numerical_results}
Input parameters are specified as follows. 
We use, according to \bib{Olive:2016xmw}, 
\begin{align}
&\alpha(0) = 1/137.035999679,\quad m_W = 80.385 \gev,\crn
&m_e = 0.5109989461 \mev, \quad m_{\mu} = 105.6583745 \mev, \quad m_{\tau} = 1776.86 \mev,\crn
&\Delta m^2_{21}=7.53\times 10^{-5} \ev^{2}, \quad \Delta m^2_{32}=2.45\times 10^{-3} \ev^2,\crn 
&\sin^2(\theta_{12})=0.307, \quad \sin^2(\theta_{13}) = 0.021, \quad \sin^2(\theta_{23}) = 0.51. 
\end{align}
The neutrino mixing matrix is assumed to be real and is 
calculated from the above mixing angles as
\bea
U^L = 
\left(
\begin{array}{ccc}
c_{12}c_{13} & s_{12}c_{13} & s_{13} \\
-s_{12}c_{23}-c_{12}s_{23}s_{13} & 
c_{12}c_{23} - s_{12}s_{23}s_{13} & 
s_{23}c_{13}\\
s_{12}s_{23} - c_{12}c_{23}s_{13} & 
-c_{12}s_{23} - s_{12}c_{23}s_{13} &
c_{23}c_{13}
\end{array}
\right),
\eea
where $c_{ij}=\cos\theta_{ij}$, $s_{ij}=\sin\theta_{ij}$ with $i,j=1,2,3$. 

\subsection{Neutrino contribution: approximate vs. exact}
\label{results_neutrino}
The approximate results calculated by neglecting the external lepton masses have been exclusively used in the literature. 
However, the justification is not totally obvious to us because the neutrino masses, which are much smaller than the lepton 
masses, are kept. We therefore present here compact formulas for the exact results (i.e. $m_1$ and $m_2$ kept) and perform 
a numerical comparison with the approximate ones. 

The SM result includes only the $W$ contribution and is given by $D_{L,R}^{\nu W}$. 
Using the formulas in \appen{appen_loops} we write the 
result in terms of scalar one-loop integrals $A_0$, $B_0$ and $C_0$, 
which are also calculated in \appen{appen_loops}. We obtain
\begin{align}
D_R^{\nu W}&=
-\fr{ieg^2}{64\pi^{2}m_W^4 t_1 (t_1-t_2)^2}\sum_{a=1}^3 U_{i_1 a}^{L \star}U_{i_2 a}^{L}
\{(t_1-t_2)(t_{a}-t_1+2)[A_{0}(m^2_W) - A_{0}(m^2_{\nu_a})]\crn
&+m^2_W \left[t_{a}^2(2t_1-t_2)-t_{a}(2t_1^2-2t_1+t_2)+t_1^2(4+t_2)-5t_1t_2-4t_1+2t_2\right]B^{(1)}_{0}\crn
&-m^2_W t_1 \left[t_{a}^2-t_{a}(t_1+t_2-1) + t_1t_2-4t_2+3t_1-2\right]B^{(2)}_{0}\crn
&-2m^4_W t_1 (t_1-t_2)(t_{a}+t_2-2t_1+2)C_0\crn
&-m^2_W t_1 (t_1-t_2) (t_{a}-t_2+2)\},\crn
D_L^{\nu W}&=
-\fr{ieg^2m_2}{64\pi^{2}m_W^4 m_1 t_{2}(t_{1}-t_{2})^2}\sum_{a=1}^3 U_{i_1 a}^{L \star}U_{i_2 a}^{L}
\{(t_{2}-t_{1})(t_{a}-t_{2}+2)[A_{0}(m^2_W) - A_{0}(m^2_{\nu_a})]\crn
&+m^2_W \left[t_{a}^2(2t_2-t_1)-t_{a}(2t_2^2-2t_2+t_1)+t_2^2(4+t_1)-5t_1t_2-4t_2+2t_1\right]B_0^{(2)}\crn
&-m^2_W t_{2} \left[t_{a}^2-t_{a} (t_{1}+t_{2}-1)+ t_{1}t_{2}-4t_{1}+3 t_{2}-2\right]B_0^{(1)}\crn
&- 2 m^4_W t_{2} (t_{2}-t_{1})(t_{a}+t_1-2t_2+2)C_0\crn
&- m^2_W t_{2} (t_{2}-t_{1}) (t_{a}-t_{1}+2)
\},
\end{align}
where $t_i=m^2_i/m^2_W$, $t_a=m^2_{\nu_a}/m^2_W$, 
$B_0^{(i)}=B_0(m^2_i,m^2_W,m^2_{\nu_a})$ with $i=1,2$ and $C_0=C_0(m^2_1,0,m^2_2,m^2_{\nu_a},m^2_W,m^2_W)$. 

\begin{figure}[h]
  \centering
  % Requires \usepackage{graphicx}
 \begin{tabular}{cc}
    % after \\: \hline or \cline{col1-col2} \cline{col3-col4} ...
  \includegraphics[width=7.9cm]{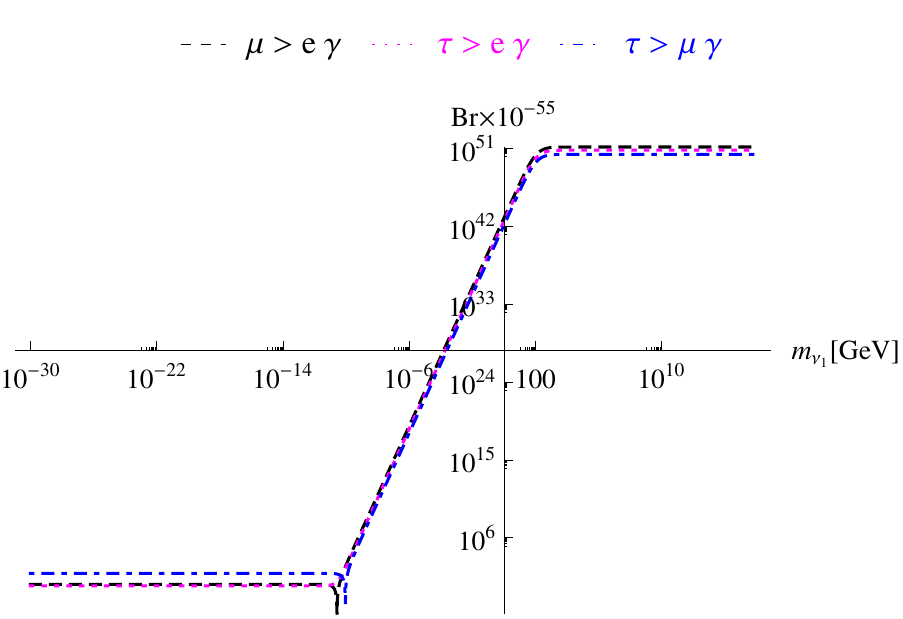}& 
  \includegraphics[width=7.9cm]{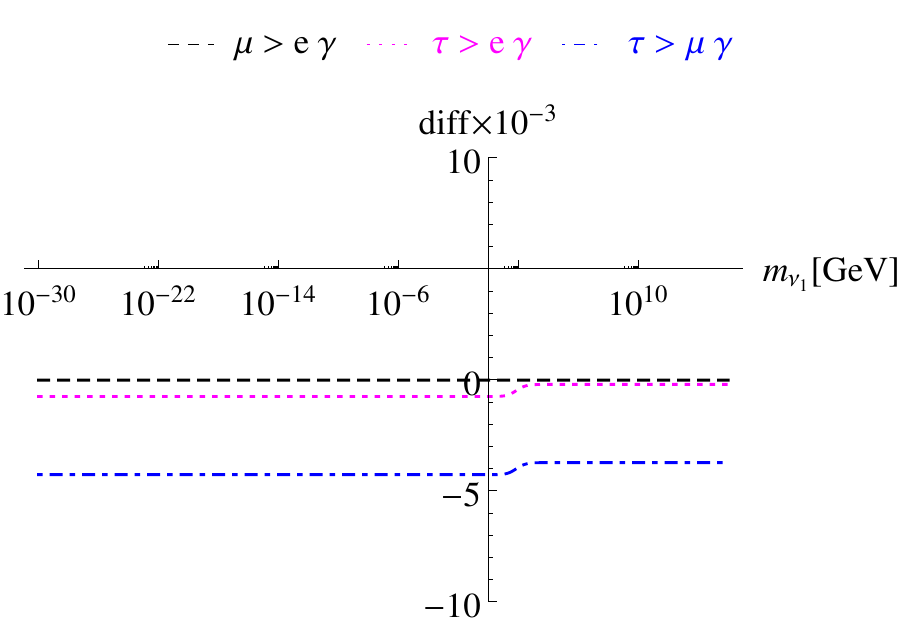}
  \end{tabular}
  \caption{Exact branching fraction (left) and difference between exact and approximate results (right) as functions of 
$m_{\nu_1}$ which we deliberately chose from very small to very large values. 
All input parameters and definitions are as in the caption of \tab{table:SM_exact_appr}. 
}\label{fig:Br_diff_SM}
\end{figure}

In the limit of $m_1=m_2=0$ we have $D_L=0$ and 
\bea
D_R^\text{appr}=\fr{ieg^2}{128\pi^{2}m^2_W}\sum_{a=1}^3 U_{i_1 a}^{L \star}U_{i_2 a}^{L}
\left[ 
\frac{10-43 t_{a}+78 t_{a}^2 -49 t_{a}^3+18 t_{a}^3 \log(t_{a}) +4 t_{a}^4}{3 (t_{a}-1)^4}
\right].
\label{eq:DR_mass_zero}
\eea
This result was first obtained in \bib{Cheng:1980tp} and has been widely used for any values of neutrino masses. 
We may wonder whether this is justified for the case of $m_{\nu_a}\ll m_1$ or $m_{\nu_a}\approx m_1$. 
This is the reason we perform a numerical comparison here between the exact result and the approximate one with $m_1=m_2=0$ 
for many values of $m_{\nu_1}$ from zero to $10^{16}\gev$. The motivation is of purely mathematical nature and 
we ignore the physical constraints on the neutrino masses here.  
The results are shown in \tab{table:SM_exact_appr} and \fig{fig:Br_diff_SM}. 
We have used \eq{eq:branching_fraction} to calculate the branching fractions for both cases. 
We see that the difference is less than permil level for $\mu \to e \gamma$ and $\tau \to e \gamma$ and is at the 
permil level for $\tau \to \mu \gamma$. This result is independent of neutrino masses. 
\begin{table}[ht!]
 \renewcommand{\arraystretch}{1.3}
  \begin{center}
   \small
%    \footnotesize
\begin{tabular}{|c|c|c|c|c|}\hline
      $m_{\nu_1}$ [GeV] & Method & $\mu\to e\gamma$ & $\tau\to e\gamma$ & $\tau\to\mu \gamma$ 
      \\\hline
      $0$ & exact Br. & $4.0969\times 10^{-55}$ & $2.6800\times 10^{-55}$ & $76.705\times 10^{-55}$ \\
          & appr. Br. & $4.0968\times 10^{-55}$ & $2.6780\times 10^{-55}$ & $76.377\times 10^{-55}$ \\
          & diff & $-2.6\times 10^{-5}$ & $-7.6\times 10^{-4}$ & $-4.3\times 10^{-3}$ \\\hline
      $10^{-13}$ & exact Br. & $4.0968\times 10^{-55}$ & $2.6801\times 10^{-55}$ & $76.705\times 10^{-55}$ \\
          & appr. Br. & $4.0967\times 10^{-55}$ & $2.6780\times 10^{-55}$ & $76.377\times 10^{-55}$ \\
          & diff & $-2.6\times 10^{-5}$ & $-7.6\times 10^{-4}$ & $-4.3\times 10^{-3}$ \\\hline
      $10^{-1}$ & exact Br. & $7.9502\times 10^{-17}$ & $3.4303\times 10^{-17}$ & $1.1400\times 10^{-17}$ \\
          & appr. Br. & $7.9500\times 10^{-17}$ & $3.4277\times 10^{-17}$ & $1.1351\times 10^{-17}$ \\
          & diff & $-2.6\times 10^{-5}$ & $-7.6\times 10^{-4}$ & $-4.3\times 10^{-3}$ \\\hline
      $10^{2}$ & exact Br. & $1.3590\times 10^{-5}$ & $0.58619\times 10^{-5}$ & $0.19481\times 10^{-5}$ \\
          & appr. Br. &$1.3590\times 10^{-5}$ & $0.58593\times 10^{-5}$ & $0.19404\times 10^{-5}$ \\
          & diff & $-2.5\times 10^{-5}$ & $-4.4\times 10^{-4}$ & $-4.0\times 10^{-3}$ \\\hline
      $10^{16}$ & exact Br. & $1.3278\times 10^{-4}$ & $0.57261\times 10^{-4}$ & $0.19030\times 10^{-4}$ \\
          & appr. Br. &$1.3278\times 10^{-4}$ & $0.57249\times 10^{-4}$ & $0.18959\times 10^{-4}$ \\
          & diff & $-2.4\times 10^{-5}$ & $-2.2\times 10^{-4}$ & $-3.7\times 10^{-3}$ \\\hline
\end{tabular}
\it{\caption[]{\small Exact (i.e. $m_1$ and $m_2$ are kept in $D_{L,R}$) and approximate (i.e. $m_1=m_2=0$) branching fractions of $l_1 \to l_2 \gamma$ at various 
hypothetical values of $m_{\nu_1}$. Other two neutrino masses are fixed at tiny values calculated using $m_{\nu_1}=0$ and the current known values of $\Delta m^2_{21}$ and $\Delta m^2_{32}$ specified in the text, namely $m_{\nu_2}\approx 8.678\times 10^{-3}\ev$ and $m_{\nu_3}\approx  5.025\times 10^{-2}\ev$. The neutrino mixing matrix is assumed being real and is 
calculated from three known mixing angles $\theta_{12}$, $\theta_{13}$ and $\theta_{23}$ as given in the text.  
For the sake of comparison we set $\text{Br}(l_1\to l_2\bar{\nu}_2\nu_1)=1$ for all three channels. The difference between exact and approximate results is defined as: $\text{diff}=(\text{appr}-\text{exact})/\text{exact}$.}
  \label{table:SM_exact_appr}}
\end{center}
\end{table} 
%%%
%---begin: plots as functions of m_{nu1}---%
%---end: plots as functions of m_{nu1}---%

We now take into account the charged Higgs contribution. There are two additional parameters $t_{v'v}$ and $m_{H^\pm}$ 
(see the $D_{L,R}^{\nu H^+}$ terms in \eq{eq:list_analytical_results}). 
We have calculated the difference between the exact and approximate results for four cases of $t_{v'v}=1/50$ or $50$ 
(we choose these exotic values so that the effect of $t_{v'v}$ is large) and $m_{H^\pm}=70$ or $700\gev$. The result 
is very similar to the SM case: the difference is below permil level for $\mu \to e \gamma$ and $\tau \to e \gamma$ 
and is at the permil level for $\tau \to \mu \gamma$. For the absolute value of $\text{Br}(\mu \to e \gamma)$ the result 
is $5\times 10^{-49}$ for $t_{v'v}=50$ and $m_{H^\pm}=70\gev$ and getting smaller for lower values of $t_{v'v}$ and/or 
higher values of $m_{H^\pm}$. 

We have a technical remark here. Due to the huge hierarchy among the neutrino, charged leptons and 
$W$ boson masses, the numerical calculation of the exact result is non-trivial because of numerical cancellation. 
To obtain the $\mu \to e \gamma$ results in \tab{table:SM_exact_appr} we have used Mathematica 9 with at least 62 precision digits for $m_{\nu_1}=10^{-13}\gev$ and about 180 precision digits 
for $m_{\nu_1}=10^{16}\gev$.

\subsection{Exotic-lepton contribution}
\label{results_exotic} 
In this numerical study we investigate the exotic-lepton contribution, to see how large the branching fractions can reach, what can be 
the dominant effects and dependence on the parameter $\beta$, $m_Y$ and $m_{H^A}$. We will also 
show the gauge-Higgs interference effects. 

In the previous section we have shown that the neutrino contribution is well below the current experimental limit. We 
will therefore neglect the neutrino contribution including interference effects with exotic leptons in the following. 
The external lepton masses will be neglected as 
justified in \sect{results_neutrino}. 

In the following we choose a benchmark point, which is a typical scenario where the $SU(3)_L$ symmetry-breaking energy scale is much larger than the SM energy scale, i.e. $m_{Y^A} \gg m_W$. If not otherwise stated, the value is chosen as
\bea
m_{Y^A} = 2\tev.
\label{eq:mY_default}
\eea
From \eq{vevang} we have 
\bea
1 + \text{ct}^2_{v'u} = \fr{m^2_{Y^A}}{m^2_W}(1 + \text{ct}^2_{v'v}),
\eea
where $\text{ct}_{v'u} = 1/t_{v'u} = u/v'$, $\text{ct}_{v'v} = 1/t_{v'v} = v/v'$. For the case of $m_{Y^A} \gg m_W$, we get
\bea
\text{ct}^2_{v'u} \approx \fr{m^2_{Y^A}}{m^2_W}(1 + \text{ct}^2_{v'v}) \gg 1.
\eea
This means that the terms proportional to $t_{v'u}$ in $D_R^{E H^A}$ in 
\eq{eq:list_analytical_results} can be safely neglected and the branching 
fractions are almost independent of $t_{v'v}$. We note that 
terms proportional to $\text{ct}_{v'u}$ are suppressed because they are 
also proportional to the external lepton masses.
We will therefore set 
$t_{v'v} = 1$ in the following. As a side note, 
for the choice of $v' = v$ there is another good justification: it makes the parameter $\rho = m^2_W/(m^2_Z \cos^2\theta_W)$ with $\theta_W$ 
being the weak-mixing angle close to unity, as pointed out in \bib{Hue:2015mna} where the same scalar potential is used. 

Other parameters related to the exotic leptons are unknown.  
We choose, as an example, the following default values for the remaining input parameters: 
\begin{align}
\beta &= 1/\sqrt{3}, \quad m_{H^{A}} = 3\tev, \crn
m_{E_1} &= 700\gev, \quad m_{E_2} = 800\gev, \quad m_{E_3} = 1\tev, \crn
\theta_{12}^{E} &= \pi/6, \quad \theta_{13}^{E} = \pi/3, \quad \theta_{23}^{E} = \pi/4. 
\label{eq:param_default}
\end{align} 
The mixing matrix $V^L$ is calculated from three mixing angles $\theta_{12}^{E}$, $\theta_{13}^{E}$, and $\theta_{23}^{E}$ 
as in the case of neutrinos.  The values 
of the exotic-lepton masses are chosen within the unitary bound of $m_{E_i} < 16m_{Y^A}$ as derived from the partial wave unitarity 
of the $E_i \bar{E}_i \to E_i \bar{E}_i$ scattering \cite{Chanowitz:1978uj}. 

A few remarks on the above default input-parameter choice are appropriate here. 
Concerning gauge bosons, the best ATLAS/CMS limits for 3-3-1 models with exotic leptons are 
summarized in \tab{table:Bound_mZ'_mY}. We note that, in almost all cases, the contributions 
from exotic leptons to the $Z'$ total width are neglected, except for the case of 
Ref.~\cite{Coutinho:2013lta} where $m_F = 1\tev$ is assumed for all 
exotic fermions. When those contributions are properly taken into account, the bound 
on $m_{Z'}$ will get weaker, because the branching fractions of $Z'\to l^+ l^-$ with $l=e,\mu$ will 
decrease. Therefore, the default choice in \eq{eq:mY_default} may be acceptable. 
However, one should keep in mind that, strictly speaking, the ATLAS/CMS bound on $m_{Z'}$ is unknown for 
our present numerical analysis, because 
it depends on the masses and electric charges of the exotic fermions (i.e. leptons and quarks) which have not been properly taken into account. 
We will therefore relax the constraint on $m_{Y^A}$, varying it from $0.5$ to $3\tev$ for some plots. 
In this context, it is noted that, using LEP II data, the authors of \bib{Buras:2013dea} obtained $m_{Z'}\gtrapprox 1\tev$ for $\beta = \pm 1/\sqrt{3}$, $2/\sqrt{3}$, leading to $m_{Y^A}\gtrapprox 0.7\tev$.    
%%%
\begin{table}[ht!]
 \renewcommand{\arraystretch}{1.3}
  \begin{center}
   \small
%    \footnotesize
\begin{tabular}{|c|c|c|c|c|c|}\hline
      $\beta$  & Data & Channel & Bound on $m_{Z'}$ & Ref. & Bound on $m_{Y^A}$ 
\\\hline
      $-2/\sqrt{3}$ & CMS8 with $20.6\fb^{-1}$ & di-muon & $\gtrapprox 3.2\tev$ & \cite{Richard:2013xfa} & $\gtrapprox 2.1\tev$  \\
\hline
      $-1/\sqrt{3}$ & CMS7{\&}8 & di-lepton & $\gtrapprox 2.5\tev$ & \cite{Coutinho:2013lta} & $\gtrapprox 2.1\tev$ \\
\hline
      $-1/\sqrt{3}$ & ATLAS8 & di-lepton & $\gtrapprox 2.89\tev$ & \cite{Salazar:2015gxa} & $\gtrapprox 2.4\tev$ \\
\hline
\end{tabular}
\it{\caption[]{\small Summary of lower bounds on $m_{Z'}$ for 3-3-1 models with exotic leptons obtained using ATLAS or CMS data 
at $7$ and $8\tev$. Exotic fermion contributions to the 
$Z'$ total width are neglected, except for Ref.~\cite{Coutinho:2013lta} where $m_F = 1\tev$ is assumed for all 
exotic fermions. In the last column we have derived the bound on $m_{Y^A}$ using the relation 
$m_{Y^A} \approx m_{Z'}\sqrt{3[1-(1+\beta^2)s^2_W]}/(2c_W)$ obtained using $v,v'\ll u$ approximation \cite{Buras:2012dp} and 
$s^2_W = 0.231$.  
}
  \label{table:Bound_mZ'_mY}}
\end{center}
\end{table}
%%%
Phenomenological constraints on the masses of exotic Higgs bosons $H^A$ 
and of the exotic leptons and their mixing angles are much more difficult and 
do not exist to the best of our knowledge. 
 
With those difficulties in mind, we decided to choose 
the above default input parameters in a fairly 
random way following a few general principles: (i) $u \gg v,v'$ (i.e. the $SU(3)_L$ breaking scale is much larger 
than that of $SU(2)_L$), (ii) the exotic leptons are heavy and satisfy the unitary bound, (iii) and their mixing angles are large. 
We note that the choice of heavy masses are in agreement with the negative results of collider searches for physics beyond the SM. 
Large mixing angles are motivated by the PMNS matrix of the neutrino sector and the fact that 
we want to have large branching fractions close to the experimental limits. 
   
In the following tables and plots, if not otherwise stated, 
the above default values are used. Differently from \sect{results_neutrino}, we will use the true values of $\text{Br}(l_1 \to l_2 \bar{\nu}_2\nu_1)$ as given in the text below \eq{eq:branching_fraction} so that one can compare the results in this section with current experimental limits.  

%%%
\begin{table}[ht!]
 \renewcommand{\arraystretch}{1.3}
  \begin{center}
   \small
%    \footnotesize
\begin{tabular}{|c|c|c|c|}\hline
      $\beta$  & $\mu\to e\gamma$ & $\tau\to e\gamma$ & $\tau\to\mu \gamma$ 
\\\hline
      $0$ & $2.51\times 10^{-13}$ & $1.94\times 10^{-14}$ & $1.57\times 10^{-16}$ \\
\hline
      $1/\sqrt{3}$ & $1.49\times 10^{-12}$ & $1.33\times 10^{-13}$ & $3.10\times 10^{-15}$ \\
\hline
      $-1/\sqrt{3}$ & $4.95\times 10^{-12}$ & $4.14\times 10^{-13}$ & $6.52\times 10^{-15}$ \\
\hline
      $\sqrt{3}$ & $2.18\times 10^{-11}$ & $1.88\times 10^{-12}$ & $3.69\times 10^{-14}$ \\
\hline
      $-\sqrt{3}$ & $3.21\times 10^{-11}$ & $2.73\times 10^{-12}$ & $4.72\times 10^{-14}$ \\
\hline
\end{tabular}
%%%
\begin{tabular}{|c|c|c|c|}\hline
      {\tiny $m_{Y^A}$[TeV]}  & $\mu\to e\gamma$ & $\tau\to e\gamma$ & $\tau\to\mu \gamma$ 
\\\hline
      $0.5$ & $1.79\times 10^{-9}$ & $1.44\times 10^{-10}$ & $1.69\times 10^{-12}$ \\
\hline
      $1$ & $6.62\times 10^{-11}$ & $5.66\times 10^{-12}$ & $1.03\times 10^{-13}$ \\
\hline
      $1.5$ & $7.38\times 10^{-12}$ & $6.49\times 10^{-13}$ & $1.41\times 10^{-14}$ \\
\hline
      $2$ & $1.49\times 10^{-12}$ & $1.33\times 10^{-13}$ & $3.10\times 10^{-15}$ \\
\hline
      $3$ & $1.64\times 10^{-13}$ & $1.47\times 10^{-14}$ & $3.61\times 10^{-16}$ \\
\hline
\end{tabular}
\it{\caption[]{\small Branching fractions of $l_1 \to l_2 \gamma$ at various 
values of $\beta$, $m_{Y^A}$. Other parameters are fixed as given in the text. 
}
  \label{table:Exotic_beta_mY}}
\end{center}
\end{table} 
%%%
%---begin: plots as functions of m_{nu1}---%
\begin{figure}[hb!]
  \centering
  % Requires \usepackage{graphicx}
 \begin{tabular}{cc}
    % after \\: \hline or \cline{col1-col2} \cline{col3-col4} ...
  \includegraphics[width=7.9cm]{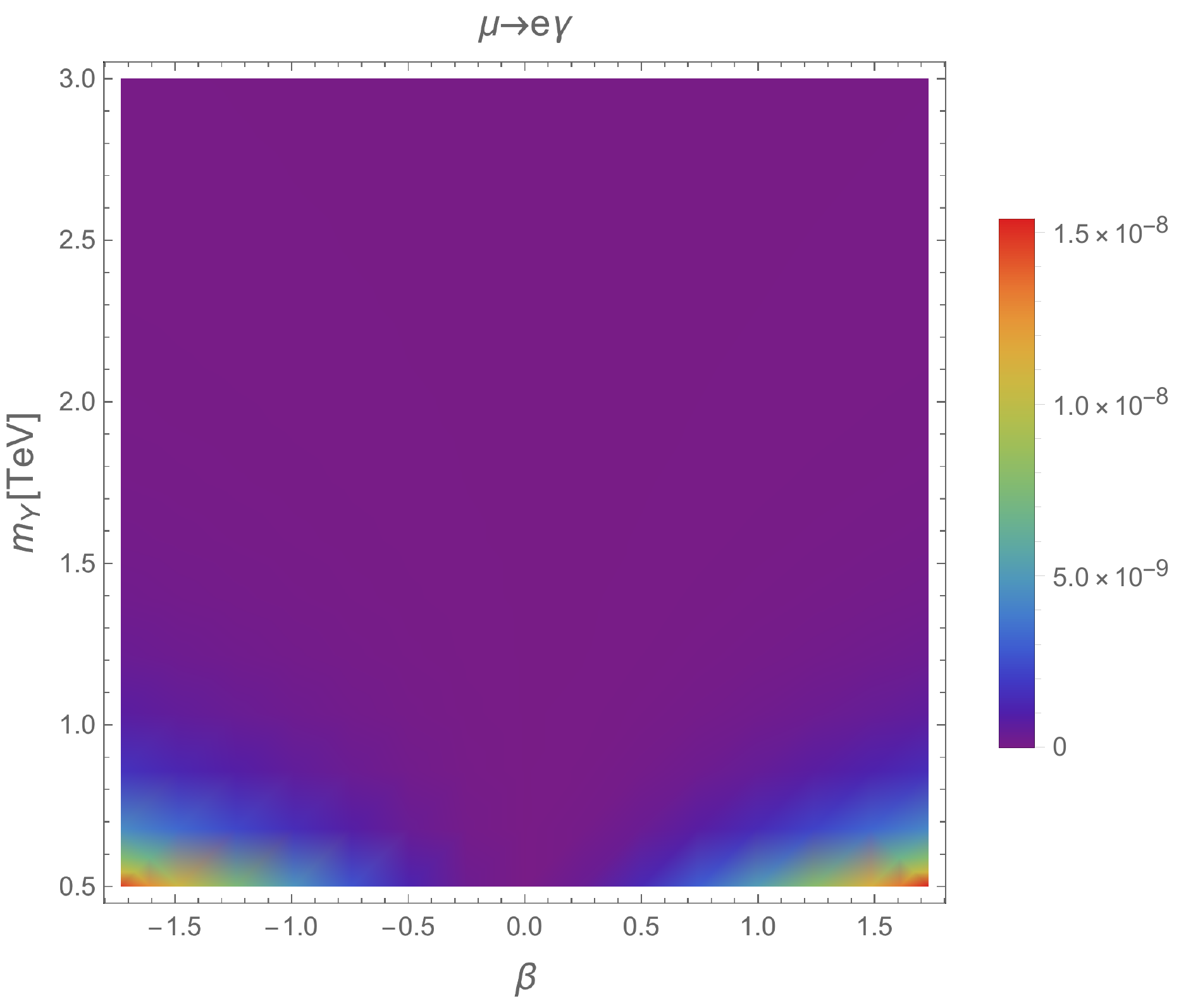}& 
  \includegraphics[width=7.9cm]{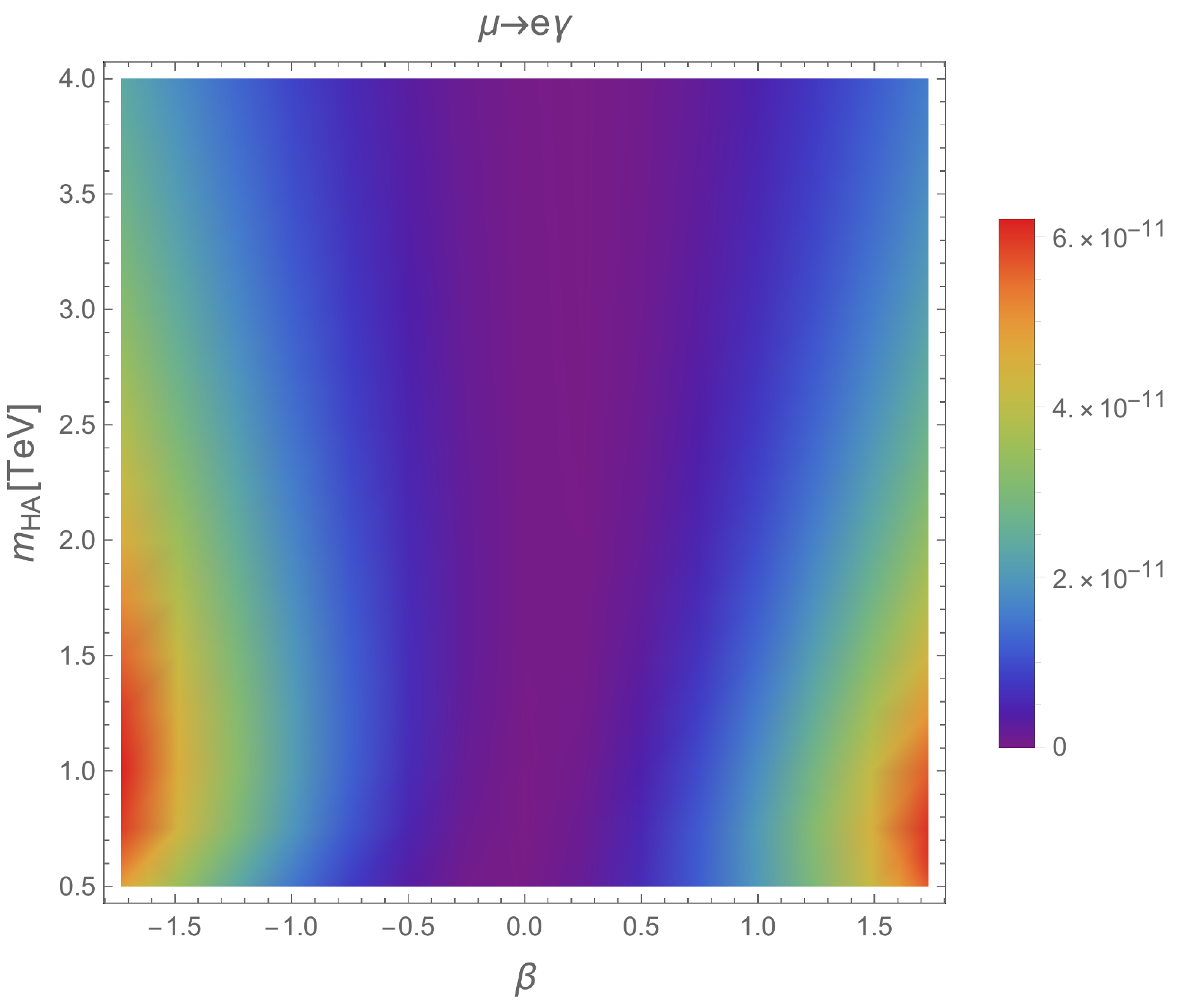}
  \end{tabular}
  \caption{Density plot of $\mu\to e\gamma$ branching fraction as function of $\beta$ and $m_Y$ (left) and 
of $\beta$ and $m_{H^A}$ (right). Other parameters are fixed as given in the text. 
}\label{fig:Density_Plots_MY_MH_beta}
\end{figure}
%---end: plots as functions of m_{nu1}---%
%---begin: plots as functions of m_{nu1}---%
\begin{figure}[hb!]
  \centering
  % Requires \usepackage{graphicx}
 \begin{tabular}{cc}
    % after \\: \hline or \cline{col1-col2} \cline{col3-col4} ...
  \includegraphics[width=7.9cm]{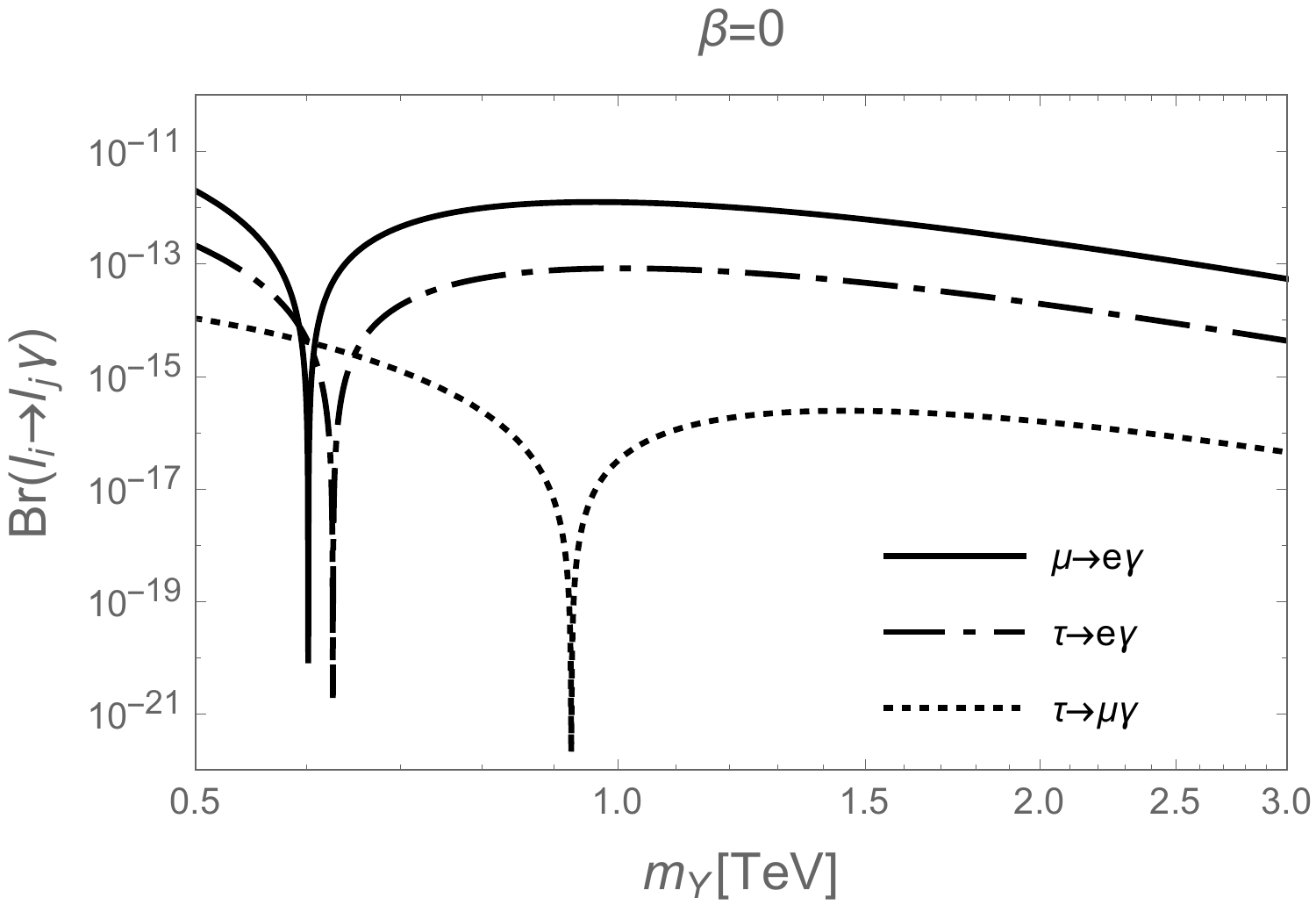}& 
  \includegraphics[width=7.9cm]{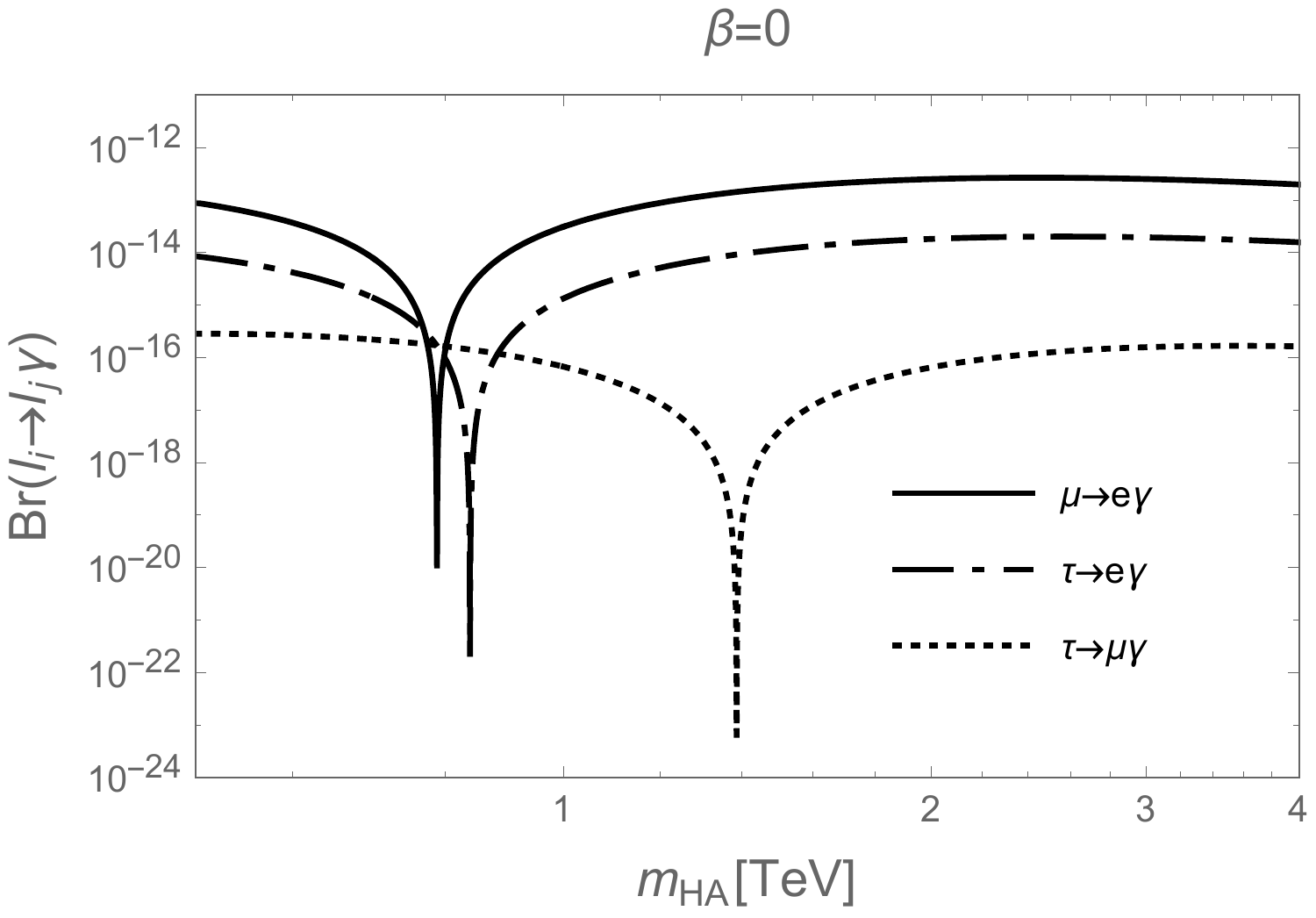}\\
%%%
  \includegraphics[width=7.9cm]{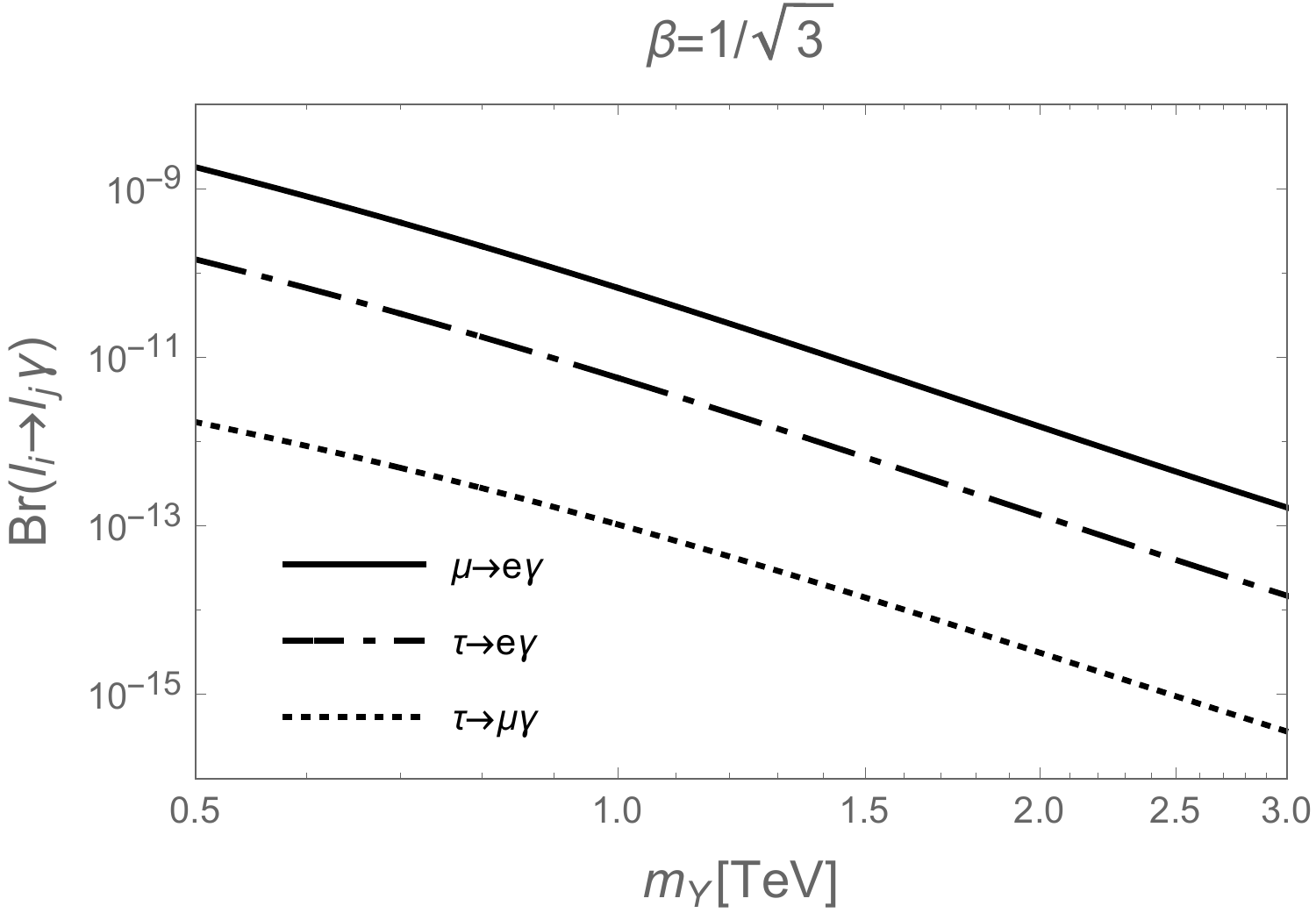}& 
  \includegraphics[width=7.9cm]{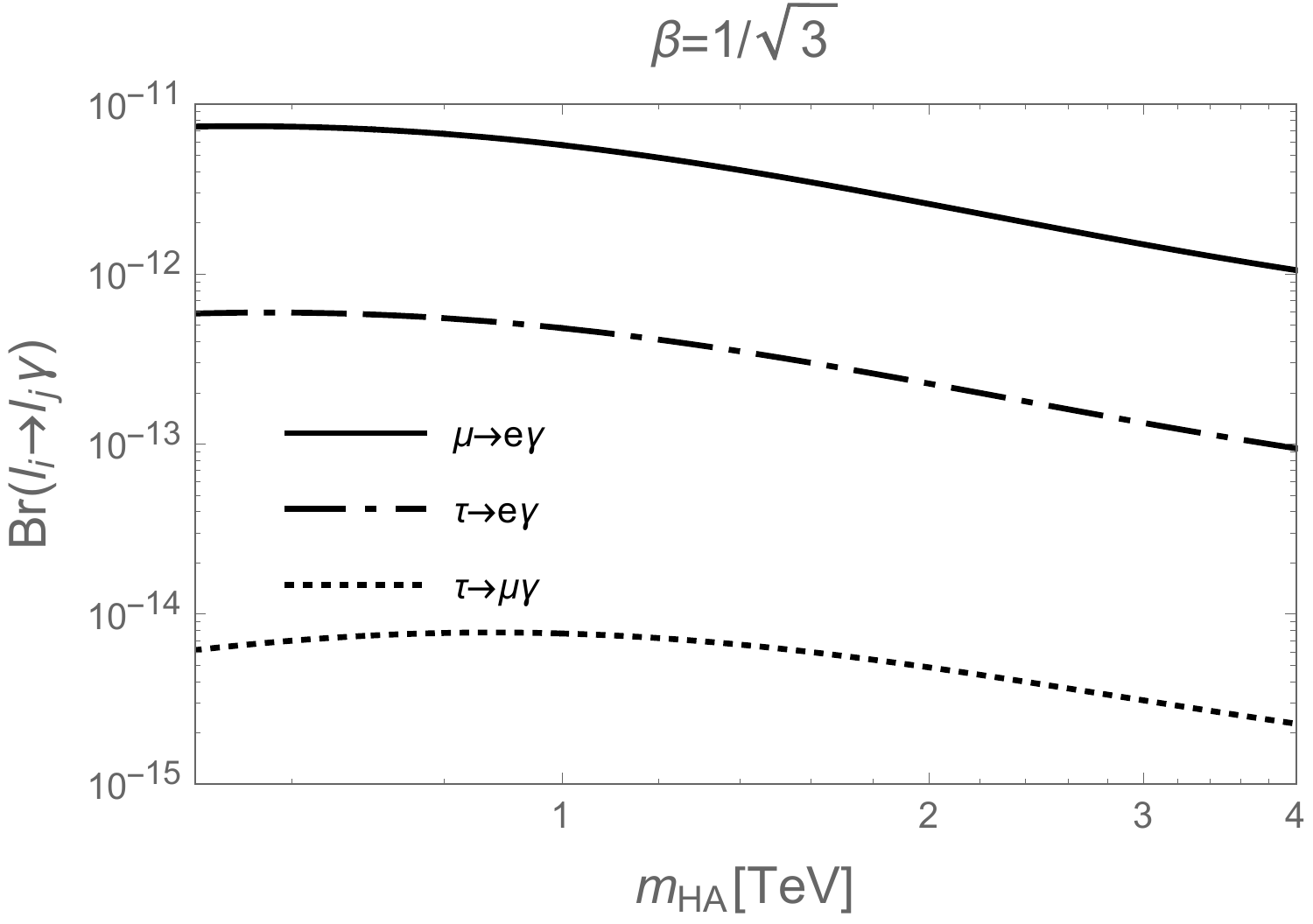}\\
%%%
  \includegraphics[width=7.9cm]{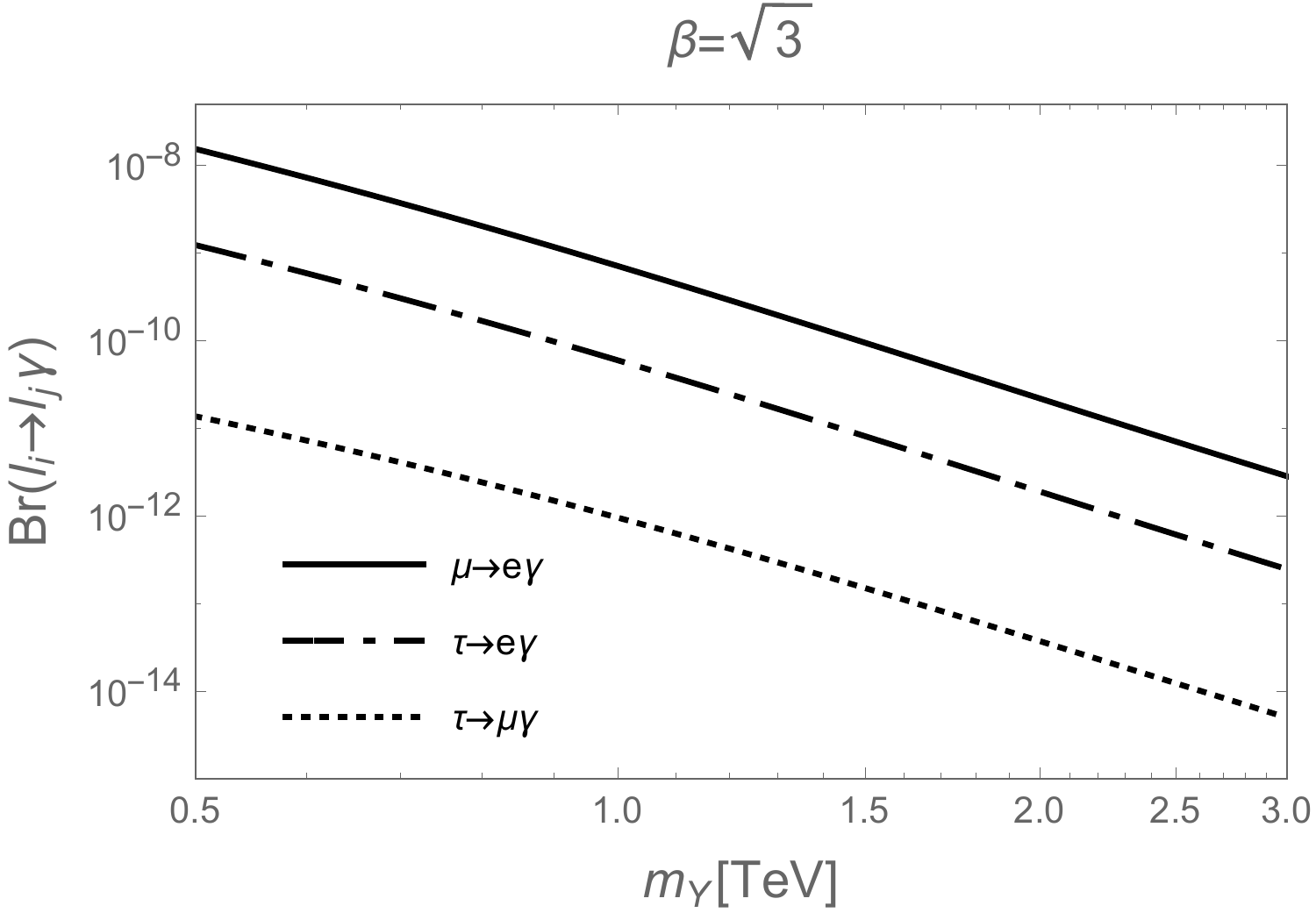}& 
  \includegraphics[width=7.9cm]{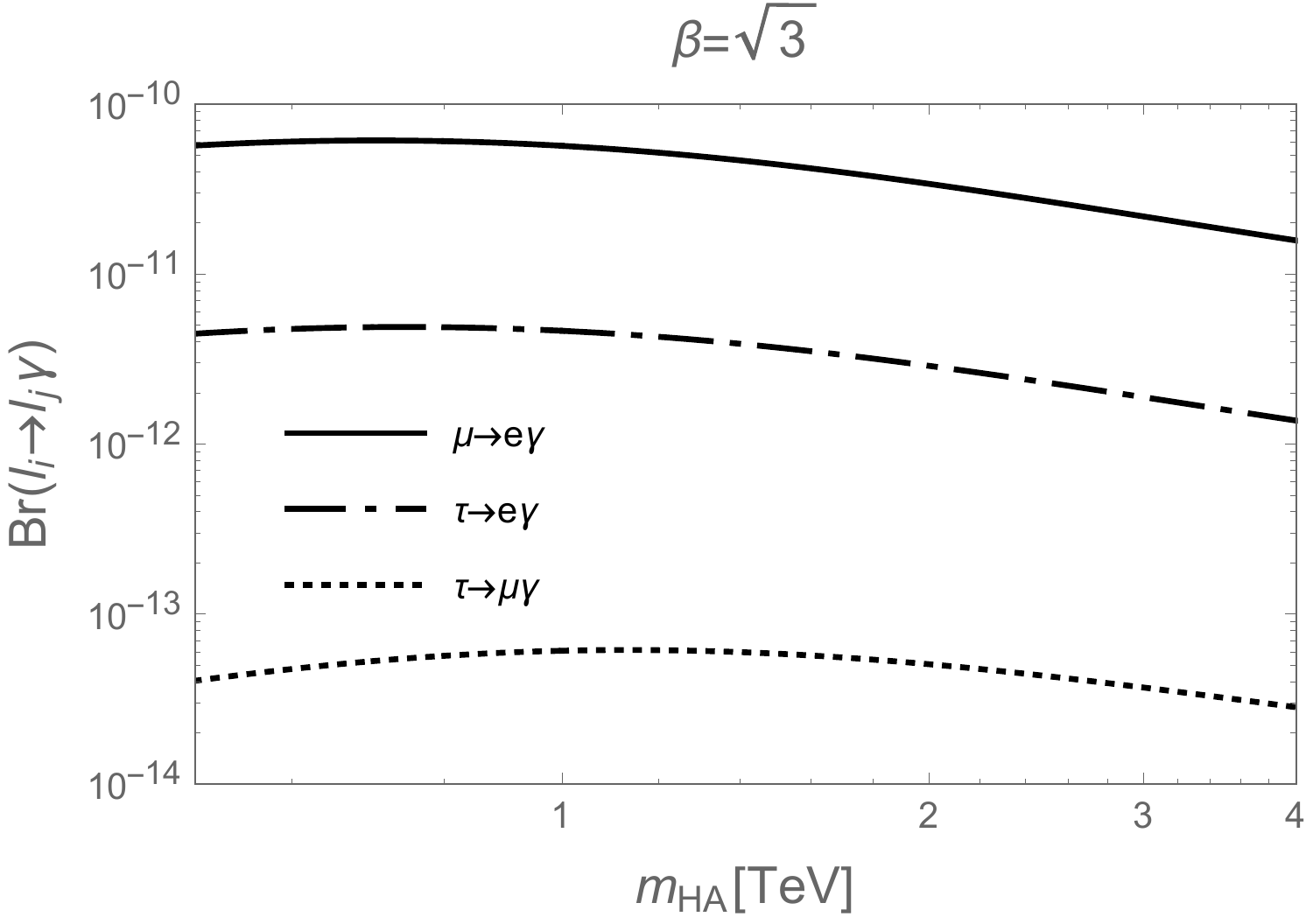}\\
  \end{tabular}
  \caption{$l_i\to l_j\gamma$ branching fractions as functions of $m_Y$ (left column) and 
of $m_{H^A}$ (right column) for various values of $\beta$: $0$ (top row), $1/\sqrt{3}$ (middle row) 
and $\sqrt{3}$ (bottom row). Other parameters are fixed as given in the text. 
}\label{fig:2D_Plots_MY_MH}
\end{figure}
%---end: plots as functions of m_{nu1}---%

In \tab{table:Exotic_beta_mY} we present the $l_1 \to l_2 \gamma$ branching fractions 
for various values of $\beta$, $m_{Y^A}$. We observe the following features: the branching fractions 
are smallest at $\beta = 0$ and increase with $|\beta|$. The results exhibit a clear asymmetry under the 
transformation of $\beta \to -\beta$, or in other words, they depend on the sign of $\beta$. 
The right table shows a strong dependence on $m_{Y^A}$. As expected, the branching fractions are large when 
$m_{Y^A}$ is small. With the choice of exotic lepton masses and mixing angles as given in \eq{eq:param_default}, 
the branching fraction is largest for $\mu \to e \gamma$ and smallest for $\tau \to \mu \gamma$. With this setup, 
we see that the branching fractions of $\tau \to e \gamma$ and $\tau \to \mu \gamma$ all satisfy the current experimental 
constraints for all values of $\beta$ and $m_{Y^A}$ in \tab{table:Exotic_beta_mY}. For the decay of $\mu \to e \gamma$, only 
the cases of $\beta = 0$ or $m_{Y^A}=3\tev$ are below the experimental limit of $4.2\times 10^{-13}$.
%%%

We now focus on the decay $\mu \to e \gamma$ and discuss two density plots to see the dependence 
on $\beta$, $m_{Y}$ and $m_{H^A}$.  
In \fig{fig:Density_Plots_MY_MH_beta} we show the density plot of $\text{Br}(\mu \to e \gamma)$ as a function 
of $\beta$ and $m_Y$ (left) and of $\beta$ and $m_{H^A}$ (right). We observe from the left plot, consistently 
with \tab{table:Exotic_beta_mY}, the branching fraction are smallest when $\beta$ is around zero or when 
$m_{Y}$ is large. From the right plot, we see a similar dependence on $\beta$, but the dependence on $m_{H^A}$ 
is much weaker than on $m_{Y}$. From those two plots, we conclude that large branching fraction occurs at large 
$|\beta|$, small $m_Y$ and small $m_{H^A}$.  

In a series of six plots in \fig{fig:2D_Plots_MY_MH} we would like to show 
again the dependence on $\beta$, $m_Y$ and $m_{H^A}$, but with 
two-dimensional plots this time and for all three decays. We see clearly that the case 
of $\beta = 0$ is special and different from the other cases of $\beta = 1/\sqrt{3}, \sqrt{3}$. 
For $\beta = 0$, the branching fractions of all three decays have a deep minimum when 
$m_Y$ or $m_{H^A}$ reach special values. The minimum positions are at low energies and are different for different decays, suggesting 
that they depend on the mixing angles. Together with \fig{fig:Density_Plots_MY_MH_beta} we conclude that 
deep minimum occurs when $|\beta|$ is small enough. This has a very important phenomenological consequence:  
for small values of $|\beta|$, branching fraction can be very small even at small values of $m_Y$ and $m_{H^A}$. 
This means that, contrary to naive expectation, there can be small values of 
$m_Y$ and $m_{H^A}$ escaping the exclusion limit obtained using the experimental constraints on $\text{Br}(l_i \to l_j \gamma)$, 
if $|\beta|$ is small enough.  

%%%

%---begin: plots as functions of m_{nu1}---%
\begin{figure}[h]
  \centering
  % Requires \usepackage{graphicx}
 \begin{tabular}{cc}
    % after \\: \hline or \cline{col1-col2} \cline{col3-col4} ...
  \includegraphics[width=7.9cm]{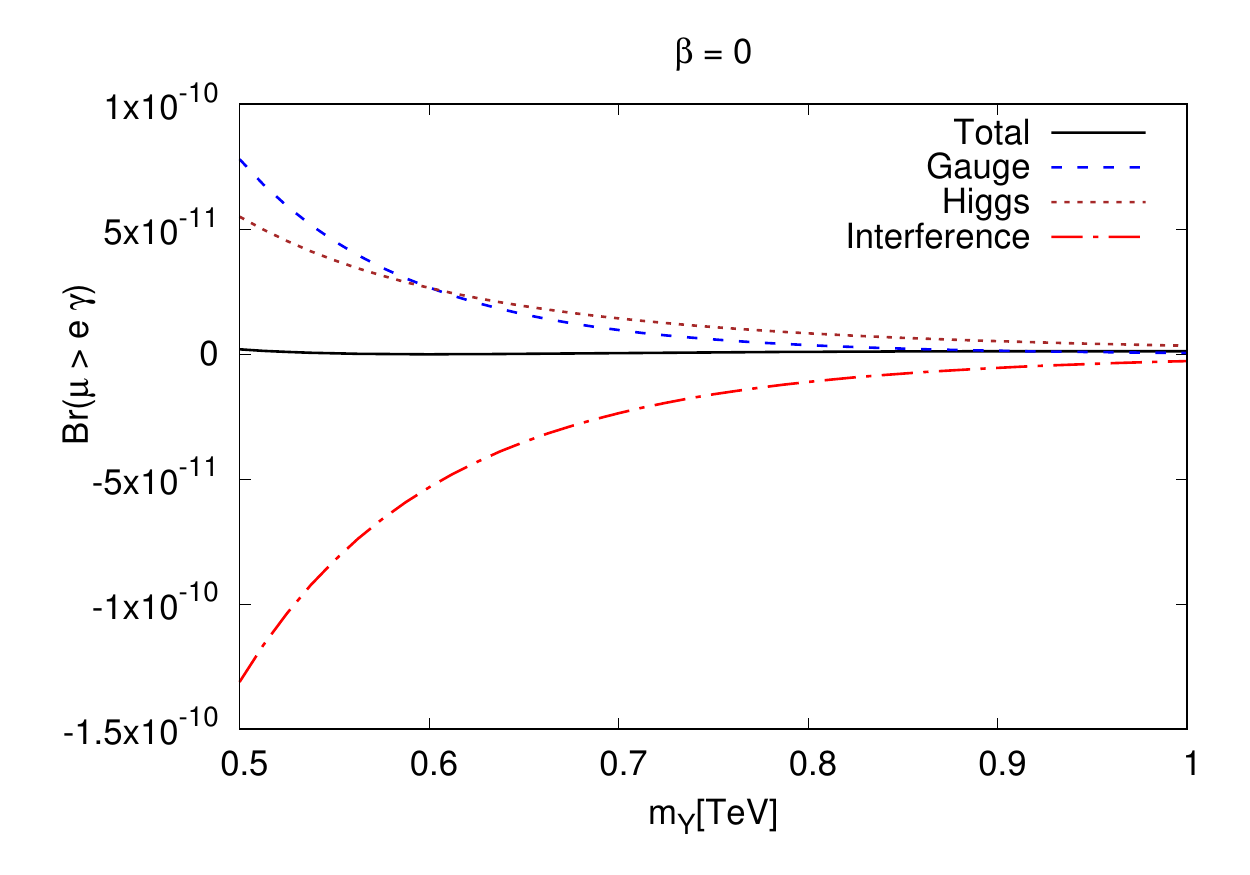}&
  \includegraphics[width=7.9cm]{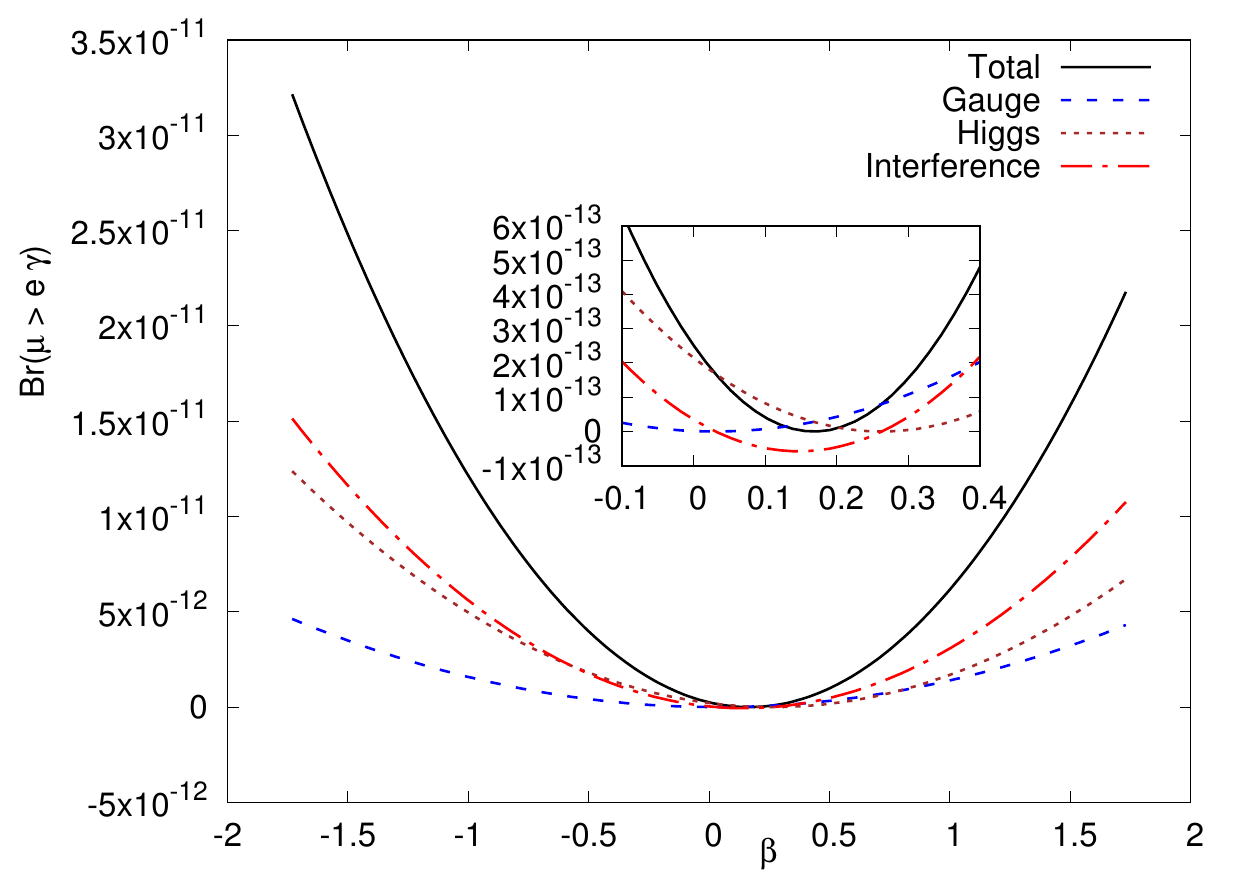}
  \end{tabular}
  \caption{Branching fraction of $\mu\to e\gamma$ as function of $m_Y$ (left) and of $\beta$ (right). 
Gauge (blue), Higgs (brown) and interference (red) contributions are also separately shown. 
Total branching fractions are the black lines.  
Other parameters are fixed as given in the text. 
}\label{fig:2D_Plots_beta}
\end{figure}
%---end: plots as functions of m_{nu1}---%

To understand the minimum occurring when $\beta$ is around zero we have to study the dependence of the 
branching fraction on $\beta$. This is shown in \fig{fig:2D_Plots_beta} (right).
On the left plot we display again the dependence on $m_Y$ for the special case of $\beta = 0$. 
This time, differently from \fig{fig:2D_Plots_MY_MH} (top-left), we focus on the low energy region of $m_Y \in [0.5,1]\tev$ 
and gauge, Higgs and interference contributions are also plotted.  
The left plot shows that the interference is strongly destructive and there is a spectacular 
cancellation between the sum of gauge and Higgs contributions and the interference term, leaving a very small 
branching ratio. The $\beta$ dependence plot also shows a negative interference effect when $\beta \in [0.035:0.26]$ for 
our default choice of input parameters. 
The insert in \fig{fig:2D_Plots_beta} (right) shows that the interference line crosses the zero branching fraction line 
when the gauge contribution (blue line) vanishes and when the Higgs term (brown line) vanishes. One should note that 
the gauge or Higgs contributions are non-negative. Overall, \fig{fig:2D_Plots_beta} shows that 
destructive interference effect tends to occur when $|\beta|$ and $m_Y$ are small.  

\section{Conclusions}
\label{conclusion}
In this paper, we have provided full and exact analytical results for the 
$l_i \to l_j \gamma$ partial decay widths for a general class of 3-3-1 models 
with exotic leptons and with arbitrary $\beta$. As a by product, we performed numerical comparisons between 
exact results (i.e. external lepton masses are kept) and approximate ones where 
$m_i = m_j = 0$. We conclude that, for either extremely light neutrinos or very heavy leptons, 
the difference between exact and approximate results is less than permil level 
for $\mu \to e \gamma$ and $\tau \to e \gamma$ and is at the 
permil level for $\tau \to \mu \gamma$. Therefore, unsurprisingly, approximation results widely 
used in the literature are excellently justified. 

Concerning the exotic lepton contribution, we found huge destructive interference between the gauge 
and Higgs contributions. This can happen when $|\beta|$ and $m_Y$ are small enough. This has an interesting consequence: 
the branching fractions can be small even for small $m_Y$. Therefore, this destructive interference mechanism 
must be taken into account when using experimental constraints on $\text{Br}(l_i \to l_j \gamma)$ to exclude parameter space. 
This in particular means that if one takes into account only the gauge contribution then the results can be completely off. 
It is likely that this destructive interference mechanism also occurs in $b \to s\gamma$ and other similar processes. 

Besides, we found that the gauge and Higgs contributions can be of similar size. Dependences on $\beta$, $m_Y$ and $m_{H^A}$ 
have been shown. We observe that the branching fractions are very sensitive to $\beta$ and $m_Y$. They also depend on $m_{H^A}$, 
but to a lesser extent. The dependence on $\beta$ is interesting: the branching fractions are largest for $|\beta|=\sqrt{3}$ and 
smallest around zero. 

\section*{Acknowledgments}
LTH would like to thank Theoretical Physics Group at IFIRSE for hospitality and supports during his stay at IFIRSE where part of this work was done. 
LDN would like to thank Jean Tran Thanh Van, Le Kim Ngoc and their team at ICISE for continuous support and creating a beautiful environment for research.  
The work of LDN has been partly supported by the German Ministry of Education and Research (BMBF) under contract
no. 05H15KHCAA. 
This research is funded by Vietnam National
Foundation for Science and Technology Development (NAFOSTED)
under Grant number 103.01-2017.29.

\appendix
\section{One loop integrals}
\label{appen_loops}
In this appendix we provide all loop functions introduced in \eq{eq:list_analytical_results}. We have
\begin{align}
h_1^{LHH}&=C_{1}([p_i^2],m_L^2,m_H^2,m_H^2)
+C_{11}(\cdots)
+C_{12}(\cdots),\crn
h_2^{LHH}&=C_{2}([p_i^2],m_L^2,m_H^2,m_H^2)
+C_{22}(\cdots)
+C_{12}(\cdots),\crn
h_3^{LHH}&=-C_{0}([p_i^2],m_L^2,m_H^2,m_H^2)
-C_{1}(\cdots)
-C_{2}(\cdots),\crn
h_4^{HLL}&=C_1([p_i^2],m_H^2,m_L^2,m_L^2)+C_{11}(\cdots)+C_{12}(\cdots),\crn
h_5^{HLL}&=C_2([p_i^2],m_H^2,m_L^2,m_L^2)+C_{22}(\cdots)+C_{12}(\cdots),\crn
h_6^{HLL}&=C_1([p_i^2],m_H^2,m_L^2,m_L^2)+C_{2}(\cdots),\crn
g_1^{LGG}&=-C_{2}([p_i^2],m_L^2,m_G^2,m_G^2)+C_{11}(\cdots)+C_{12}(\cdots),\crn
g_2^{LGG}&=C_0([p_i^2],m_L^2,m_G^2,m_G^2)+2C_{1}(\cdots)+C_{2}(\cdots)+C_{11}(\cdots)+C_{12}(\cdots),\crn
g_3^{LGG}&=C_{2}([p_i^2],m_L^2,m_G^2,m_G^2)+C_{22}(\cdots)+C_{12}(\cdots),\crn
g_4^{LGG}&=-C_{1}([p_i^2],m_L^2,m_G^2,m_G^2)+C_{22}(\cdots)+C_{12}(\cdots),\crn
g_5^{LGG}&=C_{0}([p_i^2],m_L^2,m_G^2,m_G^2)+C_{1}(\cdots)+2C_{2}(\cdots)+C_{22}(\cdots)+C_{12}(\cdots)),\crn
g_6^{LGG}&=C_{1}([p_i^2],m_L^2,m_G^2,m_G^2)+C_{11}(\cdots)+C_{12}(\cdots),\crn
g_7^{GLL}&=C_0([p_i^2],m_G^2,m_L^2,m_L^2)+2C_{1}(\cdots)+C_{2}(\cdots)+C_{11}(\cdots)+C_{12}(\cdots),\crn
g_8^{GLL}&=-C_2([p_i^2],m_G^2,m_L^2,m_L^2)+C_{11}(\cdots)+C_{12}(\cdots),\crn
g_9^{GLL}&=C_2([p_i^2],m_G^2,m_L^2,m_L^2)+C_{22}(\cdots)+C_{12}(\cdots),\crn
g_{10}^{GLL}&=C_0([p_i^2],m_G^2,m_L^2,m_L^2)+C_{1}(\cdots)+2C_{2}(\cdots)+C_{22}(\cdots)+C_{12}(\cdots),\crn
g_{11}^{GLL}&=-C_1([p_i^2],m_G^2,m_L^2,m_L^2)+C_{22}(\cdots)+C_{12}(\cdots),\crn
g_{12}^{GLL}&=C_1([p_i^2],m_G^2,m_L^2,m_L^2)+C_{11}(\cdots)+C_{12}(\cdots),
\label{eq:functions_h_g}
\end{align}
where $[p_i^2]=m_1^2,0,m_2^2$ related to external momenta and occurring in all functions, the notation $(\cdots)$ means that the same list of arguments as in the first term should be used. The masses of particles in the loop are written explicitly in the argument list and there is an 
one-to-one correspondence between those masses and the upper index of the $h_i$ ($h$ stands for Higgs) and $g_i$ ($g$ for gauge) functions. 

Using Passarino-Veltman techniques \cite{Passarino:1978jh}, the results for $C_{i\ldots}([p_i^2],m^2_F,m^2_B,m^2_B)$ read
\begin{align}
C_1&=
\frac{(m^2_1+m^2_2) \text{B}_0^{(1)}}{(m^2_1-m^2_2)^2}-\frac{2 m^2_2
   \text{B}_0^{(2)}}{(m^2_1-m^2_2)^2}-\frac{\text{B}_0^{(0)}}{m^2_1-m^2_2}
+\frac{k_2 \text{C}_0}{m^2_1-m^2_2},\crn
C_2&=
\frac{(m^2_1+m^2_2) \text{B}_0^{(2)}}{(m^2_1-m^2_2)^2}-\frac{2 m^2_1 \text{B}_0^{(1)}}{(m^2_1-m^2_2)^2}+\frac{\text{B}_0^{(0)}}{m^2_1-m^2_2}-\frac{k_1 \text{C}_0}{m^2_1-m^2_2},\crn
C_{11}&=
%\frac{\left(3 m^2_B m^4_1+4 m^2_B m^2_1 m^2_2-m^2_B m^4_2-3 m^2_F m^4_1-4 m^2_F m^2_1 m^2_2+m^2_F m^4_2-m^6_1+4 m^4_1 %m^2_2+3 m^2_1 m^4_2\right) \text{B}_0^{(1)}}{2 m^2_1 (m^2_1-m^2_2)^3}\crn
%
\frac{\left[k_1(3m^4_1-m^4_2 + 4m^2_1 m^2_2)-4m^6_1+4m^4_2m^2_1\right]\text{B}_0^{(1)}}{2 m^2_1 (m^2_1-m^2_2)^3}\crn
&-\frac{3 m^2_2 k_2 \text{B}_0^{(2)}}{(m^2_1-m^2_2)^3}-\frac{[k_1+k_2+2(m^2_2-m^2_1)]\text{B}_0^{(0)}}{2 (m^2_1-m^2_2)^2}
+\frac{\left(k_2^2+2m^2_B m^2_2\right) \text{C}_0}{(m^2_1-m^2_2)^2}\crn
&-\frac{(m^2_1+m^2_2) [\text{A}_0(m^2_B)-\text{A}_0(m^2_F)]}{2 m^2_1 (m^2_1-m^2_2)^2}
+\frac{m^2_2}{(m^2_1-m^2_2)^2},\crn
%%%
C_{22}&=
%\frac{\left(m^2_B m^4_1-4 m^2_B m^2_1 m^2_2-3 m^2_B m^4_2-m^2_F m^4_1+4 m^2_F m^2_1 m^2_2+3 m^2_F m^4_2-3 m^4_1 m^2_2-4 m^2_1 m^4_2+m^6_2%\right) \text{B}_0^{(2)}}{2 m^2_2 (m^2_1-m^2_2)^3}\crn
%
\frac{\left[k_2(-3m^4_2+m^4_1-4m^2_1m^2_2)+4m^6_2-4m^4_1m^2_2\right] \text{B}_0^{(2)}}{2 m^2_2 (m^2_1-m^2_2)^3}\crn
&+\frac{3 m^2_1 k_1 \text{B}_0^{(1)}}{(m^2_1-m^2_2)^3}-\frac{[k_1+k_2+2(m^2_1-m^2_2)]\text{B}_0^{(0)}}{2 (m^2_1-m^2_2)^2}
+\frac{\left(k_1^2+2m^2_B m^2_1\right) \text{C}_0}{(m^2_1-m^2_2)^2}\crn
&-\frac{(m^2_1+m^2_2) [\text{A}_0(m^2_B)-\text{A}_0(m^2_F)]}{2 m^2_2 (m^2_1-m^2_2)^2}
+\frac{m^2_1}{(m^2_1-m^2_2)^2},\crn
%%%
C_{12}&=
-\frac{\left[k_2(5m^2_1+m^2_2)+m^4_1-m^4_2 \right] \text{B}_0^{(1)}}{2 (m^2_1-m^2_2)^3}
+\frac{\left[k_1(5m^2_2+m^2_1)+m^4_2-m^4_1\right] \text{B}_0^{(2)}}{2 (m^2_1-m^2_2)^3}\crn
&+\frac{(2 m^2_B-2 m^2_F+m^2_1+m^2_2)\text{B}_0^{(0)}}{2 (m^2_1-m^2_2)^2}
-\frac{\left[k_1k_2+m^2_B(m^2_1+m^2_2)\right] \text{C}_0}{(m^2_1-m^2_2)^2}\crn
&+\frac{[\text{A}_0(m^2_B)-\text{A}_0(m^2_F)]}{(m^2_1-m^2_2)^2}
-\frac{m^2_1+m^2_2}{2 (m^2_1-m^2_2)^2},
\label{eq:functions_Ci}
\end{align}
where $B_0^{(0)}=B_{0}(0,m^2_B,m^2_B)$, 
$B_0^{(i)}=B_{0}(m^2_i,m^2_B,m^2_F)$, and $k_i = m^2_B - m^2_F + m^2_i$ with $i=1,2$.

The Passarino-Veltman functions in \eq{eq:functions_h_g} and \eq{eq:functions_Ci} are defined from 
the standard one-loop functions as: 
\begin{align}
&A_{0}(m^2)=\frac{(2\pi\mu)^{4-D}}{i\pi^2}\int \frac{d^D k}{k^2-m^2+i\ep},\crn
&B_{0}(p^2,m^2_F,m^2_B)=\frac{(2\pi\mu)^{4-D}}{i\pi^2}\int \frac{d^D k}{(k^2-m_F^2+i\ep)\left[(k+p)^2-m_{B}^2+i\ep\right]},\crn
&C_{0,\mu,\mu\nu}=\frac{(2\pi\mu)^{4-D}}{i\pi^2}\int \frac{d^D k (1,k_\mu,k_\mu k_\nu)}{(k^2-m_F^2+i\ep)\left[(k+p_1)^2-m_{B}^2+i\ep\right]\left[(k+p_2)^2-m_B^2+i\ep\right]},\crn
&C_\mu = p_{1\mu}C_1 +p_{2\mu}C_2,\crn
&C_{\mu\nu} = g_{\mu\nu}C_{00} + p_{1\mu}p_{1\nu}C_{11} +p_{2\mu}p_{2\nu}C_{22}+ (p_{1\mu}p_{2\nu}+p_{2\mu}p_{1\nu})C_{12},
\label{ABC_def}
\end{align}
where $\mu$ is an arbitrary mass parameter 
introduced via dimensional regularization \cite{tHooft:1972tcz}.

The scalar functions $A_0$, $B_0$, $C_0$ can be calculated using the techniques of \cite{tHooft:1978jhc}. We have 
\begin{align}
A_{0}(m^2)&= m^2\left(C_{UV} - \log(m^2) + 1\right),\crn
B_{0}(0,m^2,m^2)&=C_{UV} - \log(m^2),\crn
B_{0}(p^2,m^2_B,m^2_F)&=C_{UV} - \log(m^2_B) + 2 - \sum_{\sigma=\pm}(1-\fr{1}{x_\sigma})\log\left(1-x_\sigma\right),\crn
C_{0}(p^2_1,0,p^2_2,m^2_F,m^2_B,m^2_B)&= \fr{1}{p^2_1-p^2_2}\sum_{i=1}^{2}\sum_{\sigma=\pm}(-1)^{i}\text{Li}_2(y_{i\sigma}),
\end{align}
where $C_{UV}=2/(4-D)-\gamma_E + \log(4\pi\mu^2)$ with $\gamma_E$ being Euler's constant and $x_\sigma$ and $y_{i\sigma}$ 
are the roots of the following equations
\begin{align}
&m^2_B x^2 - (m^2_B - m^2_F + p^2)x + p^2 + i\ep = 0,\crn
&m^2_B y^2_i - (m^2_B - m^2_F + p^2_i)y_i + p^2_i + i\ep = 0.
\end{align}
For the case of $p^2_1 >0$ and $p^2_2 = 0$ we have
\begin{align}
B_{0}(0,m^2_B,m^2_F)&=C_{UV} - \log(m^2_B) + 1 + \fr{m^2_F}{m^2_B-m^2_F}\log\left(\fr{m^2_F}{m^2_B}\right),\crn
C_{0}(p^2_1,0,0,m^2_F,m^2_B,m^2_B)&= \fr{1}{p^2_1}\left[\text{Li}_2\left(1-\fr{m^2_F}{m^2_B}\right)-\sum_{\sigma=\pm}\text{Li}_2(y_{1\sigma}) \right].
\end{align}
Results for the case of $p^2_1 = p^2_2 = 0$ have been provided in \bib{Lavoura:2003xp}. 
We finally note that the $C$ functions in \eq{eq:functions_Ci} are independent of the auxiliary 
parameter $C_{UV}$, meaning that the final results are UV finite. The function $B_{0}^{(0)}$ is above given for 
the sake of completeness. The final results are independent of it. 

\section{Approximate results}
\label{appen_approx}
Here we provide results for the case of small exotic lepton masses, i.e. $m_{E_a}\ll m_Y$ and $m_{E_a}\ll m_{H^A}$. 
Furthermore, the numerical facts of $m_{\nu_a} \ll m_W$ and the approximation $m_1 = m_2 = 0$ is used as justified in \sect{results_neutrino}. 
We therefore neglect all $D_L$ here. For the neutrino case, we have
\begin{align}
D_R^{\nu W} &=-\fr{ieg^2}{32\pi^2 m_W^2}\sum_{a=1}^3 U_{i_1 a}^{L \star}U_{i_2 a}^{L}\left(\fr{m^2_{\nu_a}}{4m^2_W}\right),\crn
D_R^{\nu H^+} &=-\fr{ieg^2}{32\pi^2 m_W^2}\left(t^2_{v'v} + 6\right)\sum_{a=1}^3 U_{i_1 a}^{L \star}U_{i_2 a}^{L}\left(\fr{m^2_{\nu_a}}{12m^2_{H^+}}\right).
\label{eq:approx_neu}
\end{align}
And for the exotic lepton case
\begin{align}
D_R^{E Y} &=-\fr{ieg^2}{32\pi^2 m_Y^2}(3\sqrt{3}\beta-1)\sum_{a=1}^3 V_{i_1 a}^{L \star}V_{i_2 a}^{L}\left(\fr{m^2_{E_a}}{8m^2_Y}\right),\crn
D_R^{E H^A} &=-\fr{ieg^2}{32\pi^2 m_Y^2}\sum_{a=1}^3 V_{i_1 a}^{L \star}V_{i_2 a}^{L}\left(\fr{m^2_{E_a}}{12m^2_{H^A}}\right)\left[
\fr{\beta\sqrt{3}+1}{2}\left(t^2_{v'u}
+ 6\right)\right.\crn
&+\left.\fr{\beta\sqrt{3}-1}{2}\left(2t^2_{v'u} - 18 - 12\log{\fr{m^2_{E_a}}{m^2_{H^A}}} \right)
\right].
\label{eq:approx_exo}
\end{align}

%\bibliographystyle{h-physrev}
%\bibliography{main}

\begin{thebibliography}{10}

\bibitem{Olive:2016xmw}
Particle Data Group, C.~Patrignani {\em et~al.},
\newblock Chin. Phys. {\bf C40}, 100001 (2016).

\bibitem{Singer:1980sw}
M.~Singer, J.~W.~F. Valle, and J.~Schechter,
\newblock Phys. Rev. {\bf D22}, 738 (1980).

\bibitem{Pleitez:1992xh}
V.~Pleitez and M.~D. Tonasse,
\newblock Phys. Rev. {\bf D48}, 2353 (1993), arXiv:hep-ph/9301232.

\bibitem{Ozer:1995xi}
M.~Ozer,
\newblock Phys. Rev. {\bf D54}, 1143 (1996).

\bibitem{Diaz:2004fs}
R.~A. Diaz, R.~Martinez, and F.~Ochoa,
\newblock Phys.Rev. {\bf D72}, 035018 (2005), arXiv:hep-ph/0411263.

\bibitem{Buras:2012dp}
A.~J. Buras, F.~De~Fazio, J.~Girrbach, and M.~V. Carlucci,
\newblock JHEP {\bf 02}, 023 (2013), arXiv:1211.1237.

\bibitem{Valle:1983dk}
J.~W.~F. Valle and M.~Singer,
\newblock Phys. Rev. {\bf D28}, 540 (1983).

\bibitem{Pisano:1991ee}
F.~Pisano and V.~Pleitez,
\newblock Phys. Rev. {\bf D46}, 410 (1992), arXiv:hep-ph/9206242.

\bibitem{Foot:1992rh}
R.~Foot, O.~F. Hernandez, F.~Pisano, and V.~Pleitez,
\newblock Phys. Rev. {\bf D47}, 4158 (1993), arXiv:hep-ph/9207264.

\bibitem{Frampton:1992wt}
P.~H. Frampton,
\newblock Phys. Rev. Lett. {\bf 69}, 2889 (1992).

\bibitem{Foot:1994ym}
R.~Foot, H.~N. Long, and T.~A. Tran,
\newblock Phys. Rev. {\bf D50}, 34 (1994), arXiv:hep-ph/9402243.

\bibitem{Boucenna:2015zwa}
S.~M. Boucenna, J.~W.~F. Valle, and A.~Vicente,
\newblock Phys. Rev. {\bf D92}, 053001 (2015), arXiv:1502.07546.

\bibitem{Machado:2016jzb}
A.~C.~B. Machado, J.~Montaño, and V.~Pleitez,
\newblock (2016), arXiv:1604.08539.

\bibitem{Lindner:2016bgg}
M.~Lindner, M.~Platscher, and F.~S. Queiroz,
\newblock (2016), arXiv:1610.06587.

\bibitem{Buras:2013dea}
A.~J. Buras, F.~De~Fazio, and J.~Girrbach,
\newblock JHEP {\bf 02}, 112 (2014), arXiv:1311.6729.

\bibitem{Buras:2016dxz}
A.~J. Buras and F.~De~Fazio,
\newblock JHEP {\bf 08}, 115 (2016), arXiv:1604.02344.

\bibitem{Lavoura:2003xp}
L.~Lavoura,
\newblock Eur. Phys. J. {\bf C29}, 191 (2003), arXiv:hep-ph/0302221.

\bibitem{Griffiths:2008zz}
D.~Griffiths,
\newblock {\em {Introduction to elementary particles}}, Weinheim, Germany: Wiley-VCH (2008).

\bibitem{Vermaseren:2000nd}
J.~A.~M. Vermaseren,
\newblock (2000), arXiv:math-ph/0010025.

\bibitem{Kuipers:2012rf}
J.~Kuipers, T.~Ueda, J.~A.~M. Vermaseren, and J.~Vollinga,
\newblock Comput. Phys. Commun. {\bf 184}, 1453 (2013), arXiv:1203.6543.

\bibitem{Cheng:1980tp}
T.~P. Cheng and L.-F. Li,
\newblock Phys. Rev. Lett. {\bf 45}, 1908 (1980).

\bibitem{Hue:2015mna}
L.~T. Hue and L.~D. Ninh,
\newblock Mod. Phys. Lett. {\bf A31}, 1650062 (2016), arXiv:1510.00302.

\bibitem{Chanowitz:1978uj}
M.~S. Chanowitz, M.~A. Furman, and I.~Hinchliffe,
\newblock Phys. Lett. {\bf 78B}, 285 (1978).

\bibitem{Coutinho:2013lta}
Y.~A. Coutinho, V.~Salustino~Guimarães, and A.~A. Nepomuceno,
\newblock Phys. Rev. {\bf D87}, 115014 (2013), arXiv:1304.7907.

\bibitem{Richard:2013xfa}
F.~Richard,
\newblock (2013), arXiv:1312.2467.

\bibitem{Salazar:2015gxa}
C.~Salazar, R.~H. Benavides, W.~A. Ponce, and E.~Rojas,
\newblock JHEP {\bf 07}, 096 (2015), arXiv:1503.03519.

\bibitem{Passarino:1978jh}
G.~Passarino and M.~J.~G. Veltman,
\newblock Nucl. Phys. {\bf B160}, 151 (1979).

\bibitem{tHooft:1972tcz}
G.~'t~Hooft and M.~J.~G. Veltman,
\newblock Nucl. Phys. {\bf B44}, 189 (1972).

\bibitem{tHooft:1978jhc}
G.~'t~Hooft and M.~J.~G. Veltman,
\newblock Nucl. Phys. {\bf B153}, 365 (1979).

\end{thebibliography}

\end{document}